\documentclass[sigconf]{acmart}
\usepackage{balance}
\usepackage{multirow}
\usepackage{fixltx2e}
\usepackage{colortbl}
\newcommand{\prefixi}{\textcolor[RGB]{155, 195, 218}}
\newcommand{\prefixii}{\textcolor[RGB]{137,213,187}}
\newcommand{\prefixiii}{\textcolor[RGB]{215, 157, 164}}

\usepackage{listings}
\lstdefinelanguage{JavaScript}{
  keywords={break, case, catch, class, const, continue, debugger, default,
    delete, do, else, export, extends, finally, for, function, if, import,
    in, instanceof, let, new, return, super, switch, this, throw, try,
    typeof, var, void, while, with, yield},
  keywordstyle=\color{blue!70!black},
  ndkeywords={true, false, null, undefined},
  ndkeywordstyle=\color{purple!70!black},
  identifierstyle=\color{black},
  sensitive=true,
  comment=[l]{//},
  morecomment=[s]{/*}{*/},
  commentstyle=\color{green!50!black},
  stringstyle=\color{orange!70!black},
  morestring=[b]',
  morestring=[b]"
}
\lstdefinestyle{jsstyle}{
  language=JavaScript,
  basicstyle=\ttfamily\footnotesize,
  numbers=left,
  numberstyle=\tiny,
  frame=single,
  breaklines=true,
  breakindent=0pt,
  columns=fullflexible,
  tabsize=2,
  showstringspaces=false,
  keywordstyle=\color{blue!70!black},
  commentstyle=\color{green!50!black},
  stringstyle=\color{orange!70!black},
  xleftmargin=2em
}
\AtBeginDocument{%
  }

\setcopyright{acmlicensed}
\copyrightyear{2026}
\acmYear{2026}
\setcopyright{cc}
\setcctype{by}
\acmConference[IUI '26]{31st International Conference on Intelligent User Interfaces}{March 23--26, 2026}{Paphos, Cyprus}
\acmBooktitle{31st International Conference on Intelligent User Interfaces (IUI '26), March 23--26, 2026, Paphos, Cyprus}
\acmPrice{}
\acmDOI{10.1145/3742413.3789061}
\acmISBN{979-8-4007-1984-4/2026/03}

\acmSubmissionID{8808}

\begin{document}

\title[\textit{SCSimulator}: Visual Analytics for Supply Chain Partner Selection]{SCSimulator: An Exploratory Visual Analytics Framework for Partner Selection in Supply Chains through LLM-driven Multi-Agent Simulation}

\author{Shenghan Gao}
\orcid{0000-0003-2249-0728}
\affiliation{%
  \institution{School of Information Science and Technology, ShanghaiTech University}
  \city{Shanghai}
  \country{China}
}
\email{gaoshh1@shanghaitech.edu.cn}

\author{Junye Wang}
\authornote{Both authors contributed equally to this research.}
\orcid{0009-0001-5987-9545}
\affiliation{%
  \institution{School of Information Science and Technology, ShanghaiTech University}
  \city{Shanghai}
  \country{China}
}
\email{wangjy22022@shanghaitech.edu.cn}

\author{Junjie Xiong}
\authornotemark[1]
\orcid{0009-0004-6005-7508}
\affiliation{%
  \institution{School of Information Science and Technology, ShanghaiTech University}
  \city{Shanghai}
  \country{China}
}
\email{xiongjj2025@shanghaitech.edu.cn}

\author{Yun Jiang}
\orcid{0009-0001-1608-7919}
\affiliation{%
  \institution{Advanced Institute of Information Technology (AIIT), Peking University}
  \city{Hangzhou}
  \country{China}
}
\email{yjiang@aiit.org.cn}

\author{Yun Fang}
\orcid{0009-0008-6667-9809}
\affiliation{%
  \institution{Advanced Institute of Information Technology (AIIT), Peking University}
  \city{Hangzhou}
  \country{China}
}
\email{yfang@aiit.org.cn}

\author{Qifan Hu}
\orcid{0009-0008-5751-8704}
\affiliation{%
  \institution{Advanced Institute of Information Technology (AIIT), Peking University}
  \city{Hangzhou}
  \country{China}
}
\email{qfhu@aiit.org.cn}

\author{Baolong Liu}
\orcid{0000-0003-2628-3670}
\affiliation{%
  \institution{School of Entrepreneurship and Management, ShanghaiTech University}
  \city{Shanghai}
  \country{China}
}
\email{liubl@shanghaitech.edu.cn}

\author{Quan Li}
\authornote{Corresponding Author.}
\orcid{0000-0003-2249-0728}
\affiliation{%
  \institution{School of Information Science and Technology, ShanghaiTech University}
  \city{Shanghai}
  \country{China}
}
\email{liquan@shanghaitech.edu.cn}


\renewcommand{\shortauthors}{Gao et al.}

\begin{abstract}
Supply chains (SCs), complex networks spanning from raw material acquisition to product delivery, with enterprises as interconnected nodes, play a pivotal role in organizational success. However, optimizing SCs remains challenging, particularly in partner selection, a key bottleneck shaped by both competitive and cooperative dynamics. This challenge inherently constitutes a multi-objective dynamic game requiring a synergistic integration of Multi-Criteria Decision-Making (MCDM) and Game Theory (GT). Traditional approaches, grounded in mathematical simplifications and managerial heuristics, often fail to capture real-world intricacies and risk introducing subjective biases. Multi-agent simulation (MAS) offers promise, but prior research has largely relied on fixed, uniform agent logic, limiting practical applicability. Recent advances in Large Language Models (LLMs) create new opportunities to represent complex SC requirements and hybrid game logic. However, challenges persist in modeling dynamic SC relationships, ensuring interpretability, and balancing agent autonomy with expert control. To address these issues, we present \textit{SCSimulator}, an exploratory visual analytics framework that integrates LLM-driven MAS with human-in-the-loop collaboration for SC partner selection. \textit{SCSimulator} simulates SC evolution via adaptive network structures and enterprise behaviors, which are visualized via interpretable interfaces. By combining Chain-of-Thought (CoT) reasoning with explainable AI (XAI) techniques, the framework generates multi-faceted, transparent explanations of decision trade-offs. Users can iteratively adjust simulation settings to explore outcomes aligned with their expectations and strategic priorities. Developed through iterative co-design with SC experts and industry managers, \textit{SCSimulator} serves as a proof-of-concept, offering both methodological contributions and practical insights for future research on SC decision-making and interactive AI-driven analytics. Usage scenarios and a user study further demonstrate the system's effectiveness and usability.
\end{abstract}



\maketitle
\section{Introduction}

\par The supply chain (SC) is a complex network including activities, resources, organizations, and technologies that span the entire process from raw material acquisition to the final delivery of products or services to consumers~\cite{mentzer2001defining}. It plays a vital role in organizational success and demands close cross-departmental collaboration for efficient management. Prior research shows that a well-functioning SC not only reduces operational costs~\cite{reddy2005gaining}, but also fosters stronger upstream and downstream collaboration~\cite{spekman1988strategic} and substantially mitigates management risks~\cite{sreedevi2017uncertainty}.

\par Identifying and sustaining effective partnerships has become a central task in strengthening SC robustness~\cite{xie2022multi}, particularly within the manufacturing sector\footnote{Manufacturing enterprises typically focus on B2B transactions, yet resource and cost constraints limit the number of partnerships they can maintain.}. Specifically, the partner selection problem is a typical multi-objective dynamic game problem, which integrates the core principles from Multi-Criteria Decision-Making (MCDM) and Game Theory (GT). In practice, firms must comprehensively evaluate the capabilities of (potential) upstream and downstream stakeholders, considering factors such as operational performance and technological innovation. However, each selection does not guarantee acceptance, as counterparties are autonomous decision-makers with their own strategic considerations. Moreover, as the SC environment evolves, firms' requirements also shift, leading to changing evaluation criteria and further amplifying the complexity of partner selection.

\par Prior studies on SC have predominantly relied on mathematical modeling and statistical analysis~\cite{chen2023mathematical}. To manage the complexity of real-world dynamics and reduce computational demands, researchers often introduce simplifying mathematical assumptions. However, such assumptions lead to information loss and struggle to simultaneously capture the principles of MCDM and GT. Consequently, firms frequently rely on managerial experience to anticipate stakeholder behaviors and formulate strategies. While this approach enables comprehensive trade-off evaluation, it is labor-intensive, heavily dependent on individual expertise, and prone to subjective bias~\cite{carter2017reconceptualizing}, ultimately resulting in inconsistent decision quality. In contrast, multi-agent simulation (MAS) has emerged as a promising alternative~\cite{van1998agent,helbing2012agent}, particularly with advancements in computational capabilities. MAS represents entities as autonomous agents capable of learning and decision-making, allowing the analysis of enterprise behaviors across diverse environments while effectively integrating MCDM and GT.

\par However, traditional MAS approaches still rely heavily on mathematical models, limiting their flexibility and adaptability, despite some exploration in reinforcement learning~\cite{yan2022reinforcement,valluri2005agent}. Recent advancements in Large Language Models (LLMs) offer new opportunities due to their enhanced comprehension and reasoning capabilities. LLMs have already been applied successfully across diverse domains, including macroeconomic analysis~\cite{li2024econagent}, decision-making in warfare~\cite{hua2024warpeacewaragentlarge}, and interactions with extraterrestrial civilizations~\cite{jin2024if}. In Supply Chain Management (SCM), LLM's potential is also gaining recognition. For example, Jannelli et al.~\cite{jannelli2024agentic} proposed an LLM-driven agent framework focused on consensus-seeking within SCs, emphasizing the need for human oversight due to the current limitations of LLMs in terms of reliability and explainability. Building on these advancements, this work introduces an LLM-driven MAS that leverages LLMs' capacity to model complex requirements and hybrid game logic within SCs. Notably, given the inherent complexity of real-world SCs, such systems are not yet capable of producing definitive or fully reliable solutions. Instead, our framework should be regarded as a heuristic, exploratory tool for observing, analyzing, and interpreting the adaptive dynamics of SC networks, serving as a complement to, rather than a replacement for, rigorous analytical methods.

\par To fully harness the potential of LLM-driven MAS in assisting partner selection within SCs, three major challenges need to be addressed: 
\textbf{(1) Representing Dynamic Relationships.} Supplier-customer relationships form the backbone of SCs, making it essential for users to monitor their evolution, particularly when analyzing specific firms. In LLM-driven MAS, agents continuously adapt and restructure their relationships in response to environmental changes. However, due to the scale and complexity of these interactions, effectively visualizing such dynamic networks remains a significant challenge.
\textbf{(2) Revealing Agent Reasoning and Impact.} Simulation outputs are only meaningful when accompanied by transparent, contextual explanations. Although LLMs exhibitreasoning strong reasoning abilities, conventional textual explanations such as Chain of Thought (CoT) fall short in capturing structural dynamics or quantifying the broader impact of agent behaviors within complex environments. This limitation undermines trust, hinders insight validation, and reduces the analytical value of simulations.
\textbf{(3) Balancing Adaptive Autonomy with Expert Intervention.} In dynamic SC environments, firms must continuously adapt their focus, for example, shifting from saturated markets to emerging opportunities. While LLM-driven agents can autonomously adapt to changing contexts, expert intervention remains essential. Expert input can guide simulation trajectories, clarify complex outcomes, and ensure alignment with strategic goals, thereby enriching insight generation and maintaining practical relevance.

\par To address these challenges, we propose an interactive framework that leverages LLM-driven MAS to assist partner selection within SCs. Our approach was developed through two-phase collaborations with five domain experts, including three from industry and two from academia. In the first phase, we conducted in-depth interviews to understand current practices and pain points, which motivated our adoption of LLM-driven MAS as the foundation of the framework. In the second phase, we built a prototype and iteratively refined it through multiple rounds of expert feedback. This process allowed us to identify four key challenges and three critical findings, which together informed the formulation of seven design requirements. Guided by these insights, we developed an LLM-driven MAS pipeline capable of simulating the complex, evolving relationships among multiple firms. To improve its interpretability, we integrated LLM-based CoT reasoning with predictive models and \textit{SHapley Additive exPlanations} (SHAP)~\cite{lundberg2017unified}, providing multi-faceted explanations of agent actions and their ripple effects. On this foundation, we designed \textit{SCSimulator}, an interactive visual analytics system that enables users to track the evolution of supplier-customer relationships and actively influence simulation process with their domain expertise. We validated the efficiency and reliability of our approach through a usage scenario and a user study. In summary, this work makes the following key contributions:
\begin{itemize}
    \item We derive seven design requirements for partner selection within SCs by close collaboration with domain experts.
    \item We introduce \textit{SCSimulator}, an interactive visual analytics tool to assist decision-makers in partner selections by providing multi-faceted interpretability and rich interventions mechanisms, thereby enhancing their understanding and trust.
    \item We demonstrate the effectiveness and reliability of \textit{SCSimulator} through a usage scenario and a user study conducted with domain experts.
\end{itemize}

\section{Related Work}

\subsection{Supply Chain Analysis}

\par Supply Chain Management (SCM) has been extensively studied across multiple dimensions, including structural characteristics~\cite{wiedmer2021structural}, sustainability initiatives~\cite{sanchez2020sustainable}, digital transformation impacts~\cite{li2025digital}, and strategic management approaches~\cite{phadnis2024strategic}. Researchers have highlighted the critical role of partner selection in maintaining SC efficiency, identifying key determinants such as partner performance~\cite{ryu2009role}, R\&D investment levels~\cite{oruganti2024r}, and demand seasonality patterns~\cite{borucka2023seasonal}. MCDM have been adopted to systematically evaluate these factors, typically through weighted aggregation models that combine criteria into composite scores for partner ranking~\cite{narasimhan2006multiproduct}. Recent advancements integrate machine learning with MCDM frameworks to enhance decision robustness and reliability~\cite{ma2024decision,islam2021machine,sazvar2022sustainable}. However, these theoretical framework often overlook the strategic interdependence among firms as autonomous decision-makers, resulting in suboptimal solutions. For instance, high market-share partners may impose stringent supplier requirements that offset their apparent value, illustrating the need for strategic alignment beyond simplistic rankings~\cite{negoro2021game}. While GT models have been proposed to account for these interdependencies~\cite{leng2005game}, their practical implementation faces challenges due to the complex strategic dynamics inherent in corporate decision-making. Consequently, developing cost-effective models that balance analytical rigor with operational feasibility remains an ongoing challenge~\cite{chambers1971choose, petropoulos2024wielding}. Our work addresses this gap by developing an integrated MCDM-GT framework to optimize partner selection with scenario-specific adaptability.

\subsection{Multi-Agent Simulation}

\par Multi-agent simulation has emerged as a crucial tool in SC studies, applied in areas such as inventory control~\cite{quan2024invagent}, pricing~\cite{hou2023multi}, resource allocation~\cite{kotecha2025leveraging}, logistics management~\cite{attajer2024multi}. This approach models entities—such as manufacturers, regulators, and customers—as autonomous agents that learn and make decisions. Early studies primarily employed mathematical models, where agents optimized their decisions, allowing researchers to gain insights by analyzing their iterative behaviors and impacts~\cite{swaminathan1998modeling}. As computational power has increased, MARL has significantly advanced, enabling the simulation of more complex decision-making environments~\cite{yan2022reinforcement}. For instance, Valluri et al.~\cite{valluri2005agent} proposed a MARL framework with two distinct exploration strategies: profit-driven and quality-driven policies. Notably, some studies have incorporated hybrid simulations involving human participants~\cite{krnjaic2024scalable}. However, these approaches remain heavily dependent on mathematical assumptions, leading to inflexible, static strategies that limit their practical applications.

\par Meanwhile, LLM-driven agents have shown considerable potential due to their robust comprehension capabilities, despite the lack of direct research on partner selection within SCs. For instance, Park et al.~\cite{park2023generative} introduced \textit{Smallville}, a virtual town populated with 25 agents that interact with users through natural language, effectively simulating human behavior. Similar approaches have been explored in various domains, including macroeconomic activities~\cite{li2024econagent}, wartime decision-making~\cite{hua2024warpeacewaragentlarge}, and human-extraterrestrial interactions~\cite{jin2024if}. Additionally, Lu et al.~\cite{lu2024agentlensvisualanalysisagent} developed \textit{AgentLens}, an interactive visual analytics system that allows users to track the dynamic evolution of LLM-driven agents' events. Notably, the potential of LLMs in SCM has gained recognition, with some exploratory research already emerging. Quan et al.~\cite{quan2024invagent} introduced \textit{InvAgent}, which leverages LLM-driven agents for inventory management in virtual environments, enhancing interpretability through CoT. Wang et al.~\cite{wang2024can} developed an LLM-driven MCDM framework for supplier evaluation, demonstrating strong alignment with expert assessments. Jannelli et al.~\cite{jannelli2024agentic} proposed a series of LLM-driven agent frameworks for consensus-seeking in SCs, highlighting challenges related to reliability and explainability, particularly in complex real-world scenarios. Their work mitigates these issues through human-in-the-loop approaches.

\par Overall, current research suggests that LLMs can effectively understand the complex requirements and hybrid game logic within SCs. However, existing studies primarily focus on qualitative insights generated by LLMs, whereas decision-making in SCs requires robust quantitative reasoning. As a result, achieving fine-grained interpretability remains a  challenge. Our system integrates predictive models with SHAP-based explanations, enabling multi-faceted assessment and interpretation of LLM-generated decisions.

\subsection{Dynamic Temporal Network Visualization}

\par Temporal network visualization methods~\cite{linhares2022largenetvis} are generally categorized into two types: timeline-based and animation-based approaches~\cite{beck2017taxonomy,crnovrsanin2020staged}. Animation-based visualizations are intuitive for illustrating temporal transitions but can disrupt users' mental maps when timestamps vary significantly or when networks contain numerous elements. In contrast, timeline-based layouts effectively display multiple time steps simultaneously, but they can suffer from limited screen space as the number of nodes or timestamps increases. While both approaches support temporal analysis, timeline-based methods are typically more suitable for exploring network evolution across several discrete time points.

\par Some studies have tackled scalability challenges by visualizing large non-temporal networks containing millions of nodes~\cite{lin2013demonstrating,mi2016interactive}. However, these methods focus primarily on global structures and often lack user validation, making them unsuitable for temporal network contexts. To integrate temporal and structural analysis, other researchers have proposed summarization-based approaches, where the temporal dimension is divided into \textit{timeslices} representing network states at specific intervals. While such techniques perform well for discrete-time networks, continuous-time networks~\cite{arleo2021multilevel} require specialized solutions such as \textit{DynNoSlice}~\cite{simonetto2017drawing,simonetto2018event} and \textit{MultiDynNoS}~\cite{arleo2021multilevel} to represent dynamic events. Despite these advances, most temporal network visualization techniques still struggle to highlight how specific nodes—and their associated connections—evolve over time.

\par To overcome this limitation, dynamic egocentric network visualization has emerged as a specialized approach focusing on the evolving relationships between a central entity (the ``ego'') and its connected entities (``alters'')~\cite{wu2015egoslider,li2017visual}. Unlike global network visualizations, egocentric methods emphasize the local subnetwork surrounding the focal node, aligning well with our use case where each decision-maker acts as an independent node within SCs rather than as a macro-level regulator. For visualizing such dynamic ego-networks, both animation and timeline-based techniques have been used~\cite{beck2017taxonomy}; however, given the high cognitive load associated with animations in analytical tasks~\cite{robertson2008effectiveness}, we focus on timeline-based visualization as a more effective strategy for revealing the processes within LLM-driven MAS.

\par Existing research on dynamic egocentric visualization can be broadly divided into two categories: node-link–based methods~\cite{liu2017egocomp,he2016mena} and line-based methods~\cite{fu2021dyegovis,zhao2016egocentric,kuo2024spreadline}. Node-link–based methods focus on structural representations at individual timestamps. For instance, \textit{MENA}~\cite{he2016mena} visualizes discrete-time network evolution, while Farrugia et al.~\cite{farrugia2011exploring} employed a radial layout with time encoded as radius. Other works use visual metaphors—such as pollen dispersal~\cite{barker2021pollen}, tree diagrams~\cite{sallaberry2016contact}, or hexagonal grids~\cite{burdziej2019using}—to depict the ego's influence on its alters. However, these designs often encounter scalability issues, resulting in visual clutter~\cite{yoghourdjian2020scalability}. Line-based approaches mitigate these issues by enabling clearer temporal tracing of nodes. \textit{EgoSlider}~\cite{wu2015egoslider} employs vertical positioning and color encoding to represent changes in ego-network structure over time. \textit{EgoLines}~\cite{zhao2016egocentric} uses a ``subway map'' metaphor to trace nodes through the timeline, while \textit{DyEgoVis}~\cite{fu2021dyegovis} adopts a diagonal force-directed layout to balance readability and topology preservation. Similarly, \textit{SpreadLines}~\cite{kuo2024spreadline} integrates positional and color encodings within a storyline framework to concisely depict topological and contextual information.

\par Building on these insights, we adopt a timeline-based strategy to track SC evolution. Following the ``overview first, zoom and filter, then details-on-demand'' principle, we designed two linked visualization views. The \textit{Global View} provides a macro-level perspective, displaying SC evolution as a series of consistent 2D embeddings that connect nodes across time. The \textit{Focus View}, inspired by \textit{EgoLines} and \textit{SpreadLines}, employs a line-based egocentric visualization enriched with a ``berry orchard'' metaphor to highlight bilateral relationships and dynamic interactions within the SC.

\section{Formative Study}
\par To inform our design process, we conducted an iterative, feedback–driven study with five domain experts via periodic semi-structured interviews. This section presents our methodology and highlights the key insights derived from their input.

\subsection{Participant \& Procedure}
\par We collaborated with five experts (\textbf{E1-E5}) with diverse backgrounds. The panel included three senior managers from manufacturing enterprises (\textbf{E1-E3}), where \textbf{E2} serves as a supplier to \textbf{E1}, and \textbf{E3} operates independently. Additionally, two academic researchers specializing in SCM (\textbf{E4-E5}) with prior publications in the field contributed from a partner university.

\par The formative study was structured into two sequential phases. In \textbf{Phase 1}, we conducted expert consultations to examine current practices and identify key challenges, which motivated our adoption of an LLM-driven MAS framework. In \textbf{Phase 2}, we created a prototype and iteratively refined it through multiple rounds of expert feedback, resulting in three core findings. Insights from both phases informed the derivation of seven concrete design requirements. 

\subsection{Phase 1: Current Practices and Challenges}

\par Experts shared valuable insights into their workflows for partner selection within SCs. In industry, the process typically begins with gathering information on (potential) suppliers, (potential) customers, and competitors. Managers then develop multiple strategic alternatives based on specific objectives, such as market expansion. These strategies are later synthesized into a final decision, which is then implemented. Throughout this process, knowledge derived from MCDM and GT is often implicitly integrated into managerial thinking and discussions.
In contrast, researchers approach partner selection through two main methods: constructing mathematical models based on theoretical assumptions and conducting statistical analyses of empirical data. In both approaches, MCDM and GT serve as flexible frameworks that guide the modeling or analytical structure. Based on these insights, we identified four key challenges, organized by stakeholder type: \prefixi{\textbf{[Enterprise]}}, \prefixii{\textbf{[Researcher]}} and \prefixiii{\textbf{[Bilateral]}}.

\par \prefixi{\textbf{[Enterprise]}} \textbf{C1. The disruptive impact of subjective bias on decision quality.} In industry, decision-making is often conducted by small leadership groups in meetings. \textbf{E1-E3} expressed concerns that this approach relies heavily on individual expertise, which can introduce subjective bias into the process. As \textbf{E2} pointed out, ``\textit{You know how we sometimes get stuck in that `I think this is okay' trap? Like, when we're signing a contract, we might accidentally assume the other side shares the same perspective—and end up misjudging their priorities completely.}''

\par \prefixii{\textbf{[Researcher]}} \textbf{C2. Information loss due to over-abstraction.} In academia, simplification is commonly employed to reduce computational complexity. However, researchers have raised concerns that this approach can lead to information loss, as specific requirements and contextual factors are abstracted into symbols and formulas. As \textbf{E4} pointed out, ``\textit{To make things clearer, we often simplify specific requirements and external factors into mathematical formulas. Yet, this kind of simplification can strip away important details, making it hard to capture real-world complexities and limiting the broader applicability of the findings.}''

\par \prefixii{\textbf{[Researcher]}} \textbf{C3. Mechanistic alignment in decision pathways.} Current research methodologies often overlook the diversity of decision criteria within SCs. While some GT models, such as the \textit{signaling game}~\cite{banks1987equilibrium} account for different roles like sender and receiver, researchers have strengthened that homogenizing decision criteria diminishes the reliability of the outcomes. As \textbf{E5} explained, ``\textit{In my research, I typically use a single model for all agents. But in reality, decision-making logic varies significantly because companies have different operational approaches, and decision-makers each their own unique styles.}''

\par \prefixiii{\textbf{[Bilateral]}} \textbf{C4. Divergent goals rooted in stakeholder roles.} We observed distinct evaluation priorities among participants, shaped by their companies' strategic goals or research objectives. For instance, \textbf{E1}, representing a startup aiming to disrupt the industry, prioritized technological advancement, focusing on partners capable of helping overcome technical barriers and accelerating product innovation. In contrast, \textbf{E3}, from a well-established company with a mature SC system, favored partnerships with large-scale enterprises due to their lower risk of default. As \textbf{E3} explained, ``\textit{Smaller companies aren't necessarily worse than well-known firms in terms of product performance or pricing. But in general, they have weaker risk resistance, making them more likely to default or even go bankrupt.}'' These differences highlight the difficulty in supporting users' various expectations.

\par Notably, \textbf{E4} and \textbf{E5} highlighted MAS as a promising approach to integrate MCDM and GT. While MAS has not yet become mainstream due to its technical complexity, both our literature review and expert discussions support its adoption as a foundational design technique. However, further discussion revealed that traditional MAS approaches, whether based on formal models or reinforcement learning, continue to face challenges related to \textbf{C2} and \textbf{C3}, as they still rely on mathematical model and assumptions. This led us to explore LLM-driven MAS as a potential solution to these challenges, leveraging LLM's ability to comprehend nuanced decision contexts, dynamically adapt agent behavior, and facilitate hybrid reasoning across diverse stakeholder priorities.

\subsection{Phase 2: Findings in LLM-driven MAS}

\par To further exploit the potential of LLM-driven MAS for partner selection in SCs, we developed a prototype and conducted multiple iterative co-design sessions with domain experts. Each iteration centered on a distinct aspect of the system's capabilities, ranging from visualizing simulation outcomes and generating interpretable explanations to enable expert-driven guidance during simulations. This co-evolutionary process not only refined our design but also deepened our collective understanding of how experts interact with autonomous agents in dynamic simulation contexts. 
\par In the early stage, we explored how to effectively visualize the outcomes of the LLM-driven MAS. We experimented with various forms of temporal network visualizations, including animation-based force-directed layouts and timeline-oriented ego-centric views. However, in an LLM-driven MAS, firms represented as agents, continuously adapt and reconfigure their SC relationships. This dynamic restructuring presents challenges for experts in tracking how the overall topology evolves over time, particularly when comparing multiple focal firms simultaneously. Through this exploration, we identified \textbf{F1. From observing network changes to supporting tracking of evolving structural complexity}, highlighting the need for a coherent understanding of evolving network structures amidst frequent and interdependent changes.

\par As the collaboration evolved, we realized that visualizing simulation results alone was insufficient. Experts consistently expressed the need to understand the impacts of decisions, rather than just observing their outcomes. Initially, we relied on a prompt-based paradigm for explanation, as it was cost-effective and required no additional training data. However, these textual explanations primarily conveyed the reasoning behind decisions, without providing a quantitative view of how forming or dissolving a partnership affected both parties. In response to expert feedback, we integrated traditional analytical models to explicitly examine agents' decision processes and quantify their mutual impacts. Through this process, we identified \textbf{F2. From observation to understanding through explanation}, highlighting experts' need for both qualitative and quantitative interpretability to better comprehend how agent decisions influence network dynamics.

\par In later iterations, we noticed that experts' analytical focus shifted dynamically as the simulation unfolded. Initially guided by predefined assumptions or hypotheses, experts often adjusted their attention after encountering unexpected outcomes. This insight led us to reconceptualize the simulation process—not as a fixed, closed-loop execution, but as an interactive exploration that adapts to emerging interests. Consequently, we identified \textbf{F3. From predefined focus to adaptive expert-driven exploration}, highlighting the need to support experts in redirecting the course of simulations in response to evolving objectives, while also documenting these shifts in analytical focus.

\subsection{Design Requirements}
\par Drawing from the identified challenges and contextual findings, we derived the following design requirements.

\par \textbf{DR1. Rapid scenario configuration.} To address diverse simulation needs (\textbf{C4}), the system should support flexible scenario construction. It should accommodate various forms of unstructured input, such as firm profiles, scenario descriptions, and simulation parameters such as number of simulation rounds.

\par \textbf{DR2. Context-aware autonomous decision-making.} Agents should possess the ability to independently analyze and make decisions, rather than relying solely on human guidance (\textbf{C1}). This decision-making process should be grounded in scenario-specific analysis (\textbf{C2}), ensuring that the system specifies how agents receive, process, and act on information to accurately assess the SC and produce rational behaviors.

\par \textbf{DR3. Heterogeneous agent behavior.} To avoid the homogenized decision-making tendencies common in traditional MAS (\textbf{C3}), agents must have independent decision-making logic. While \textbf{DR2} emphasizes autonomy in relation to humans, \textbf{DR3} highlights autonomy among agents themselves. The system should clearly distinguish between global rules and agent-specific behaviors to allow for diverse decision-making styles.

\par \textbf{DR4. Reveal agent decision logic and its impact.} 
Experts highlighted that the value of a simulation lies not only in the insights it generates but also in understanding the rationale behind those insights (\textbf{F2}). Although LLMs can explain decisions through textual reasoning, such as CoT, these narratives often fall short of illustrating how each action affects other agents or the broader network. Therefore, the system should combine LLM reasoning with quantitative attribution mechanisms to provide clear, interpretable explanations of agent behaviors and their systemic impacts.

\par \textbf{DR5. Visualize supply chain interactions.} 
The dynamic nature of LLM-driven MAS produces a high volume of temporal changes in SC relationships (\textbf{F1}). The system should offer scalable and intuitive visualizations that allow users to trace the evolution of network structures over time, providing fine-grained insights into interactions between agents.

\par \textbf{DR6. Facilitate expert involvement.} 
Given the diverse priorities of experts (\textbf{C4}) and their evolving interests during simulations (\textbf{F3}), the system must support flexible expert involvement. This includes adjusting agent priorities, modifying decision criteria, and customizing specific simulation interactions to ensure that simulations remain aligned with domain-specific goals and evolving analytical needs.

\par \textbf{DR7. Record and trace simulation paths.}
Since experts may continuously adjust the direction of simulations during the process (\textbf{F3}), the system should record the evolution of simulation paths and support branching. This will allow experts to explore alternative trajectories based on emerging interests, providing a flexible and dynamic approach to simulation exploration.

\section{SCSimulator}

\par
\begin{figure}[h]
    \centering
    \includegraphics[width=\linewidth]{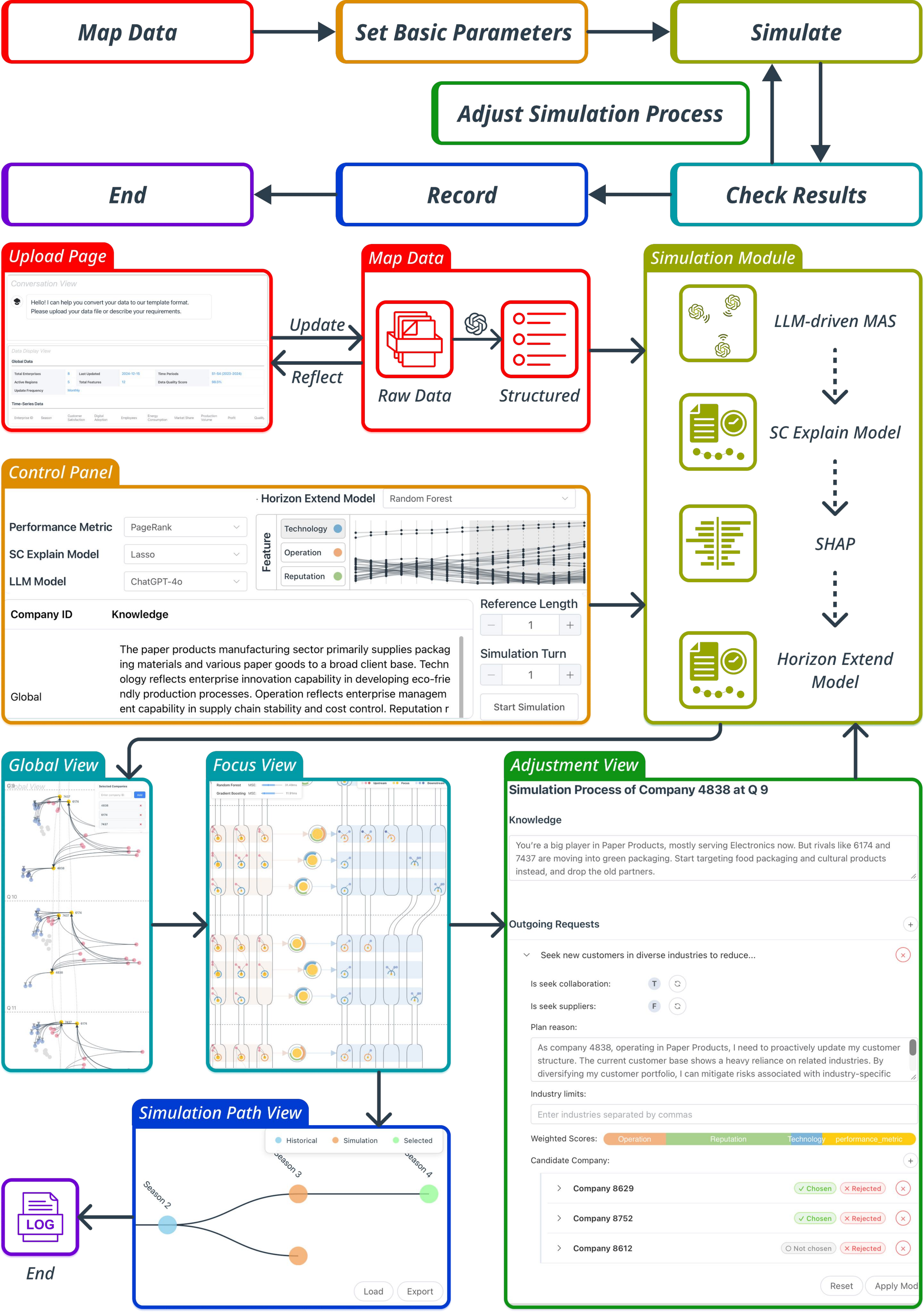}
    \caption{Framework of \textit{SCSiumulator}. It integrates an LLM-driven multi-agent system with interactive visual analytics. Users map raw data, configure parameters, run simulations, and iteratively refine agent behaviors through a human-in-the-loop workflow.} 
    \label{fig:pipeLine}
\end{figure}

 As illustrated in \autoref{fig:pipeLine}, \textit{SCSimulator} integrates a human-in-the-loop collaboration mechanism to support partner selection within SC through an LLM-driven MAS framework. In the initial \textit{Map Data} stage, a conversational LLM agent assists users from diverse backgrounds in mapping their raw data to the simulator's structured templates via \textit{Upload Page}(\textbf{DR1}). Users then configure simulation parameters via the \textit{Control Panel} (\textbf{DR1}). The simulation module runs autonomous MAS based on these input and settings (\textbf{DR2,3}). It then analyzes the resulting scenarios to reveal network dynamics (\textbf{DR4}). Users can explore it from a global perspective via the \textit{Global View} (\textbf{DR5}) and analyze node-level mechanisms through the \textit{Focus View} (\textbf{DR4, 5}). The \textit{Adjustment View} enables users to review detailed simulation procedures and interactively guide agent behaviors (\textbf{DR6}), while the evolving simulation trajectories are recorded and visualized in the \textit{Simulation Path View} to ensure traceability (\textbf{DR7)}. Users may further refine their exploration or export simulation logs for extended analysis.
\par The framework comprises a backend engine and a frontend interface with interactive visualizations, which are detailed in the following sections.

\subsection{Backend Engine}

\subsubsection{Modeling Companies}

\par Within \textit{SCSimulator}, the SC is conceptualized through three complementary dimensions, \textbf{Network}, \textbf{Feature}, and \textbf{Knowledge}. This tri-dimensional representation enables \textit{SCSimulator} to construct comprehensive firm profiles and effectively capture the dynamic of inter-firm relationships (\textbf{DR2}).

\par \textbf{Network} represents the temporal linkages between suppliers and customers, capturing the dynamic evolution of inter-firm relationships over time.

\par \textbf{Feature} characterizes company profiles, incorporating both non-temporal attributes (e.g., industry type) and temporal attributes (e.g., innovation indicators).

\par \textbf{Knowledge} provides LLM-driven agents with the contextual foundation for strategic reasoning. It includes global knowledge, such as implicit interpretations of firm features, as well as firm-specific knowledge that informs strategic objectives and guides company-level decision-making.

\par For \textbf{Company Performance}, we adopt network-based metrics, such as the \textit{number of collaborators} and \textit{PageRank}~\cite{page1999pagerank}, to quantify the relative importance of each company within the overall SC network, following established management research~\cite{kim2018supplier} and \textbf{E4}'s suggestions. Users may select specific metrics according to their analytical objectives.


\subsubsection{LLM-driven Multi-Agent Simulation}

\label{sec:agent-framework}

\par
\begin{figure}[h]
  \centering
  \includegraphics[width=\linewidth]{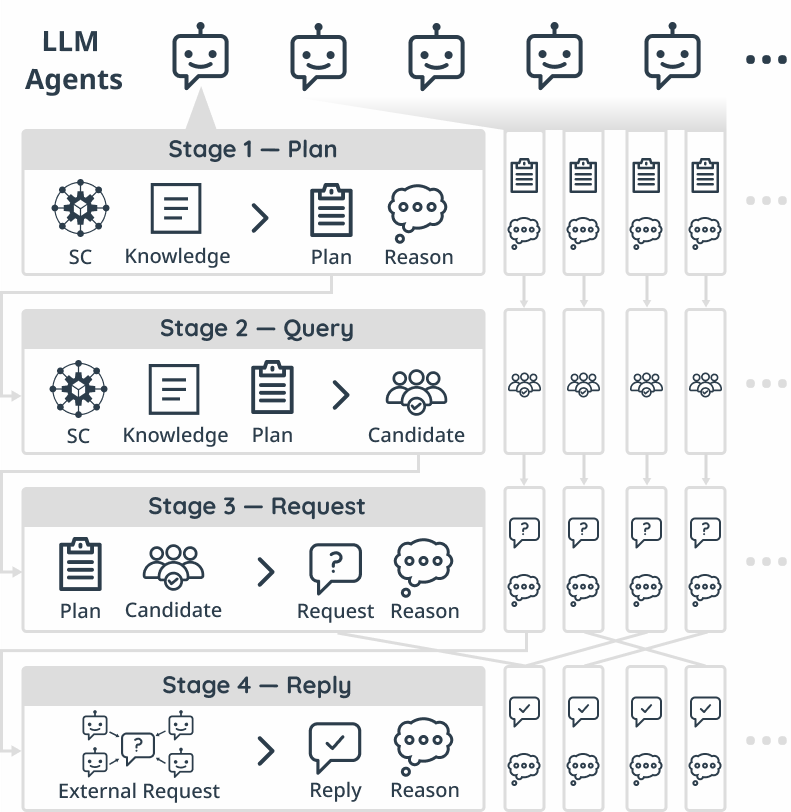}
    \caption{Pipeline of LLM-driven MAS. Each firm is modeled as an autonomous LLM agent that collaboratively plans, queries, requests, and replies to others. The four-stage request–response cycle simulates partner-seeking behaviors and the evolving dynamics of SC networks.}
  \label{fig:MASPipeLine}
\end{figure}

 To simulate the evolution of SCs and the strategic interactions among firms, we model each company as an autonomous agent operating within a partnership-seeking framework (\textbf{DR2,3}). This framework follows a ``request-response'' cycle, aligned with established methodologies in business partner selection~\cite{lau2001development,schoop2003negoisst}. While LLMs possess strong reasoning and comprehensive capabilities, directly encoding an entire SC into a single prompt is infeasible due to its scale and complexity. To address this, we propose a four-stage simulation framework comprising \textbf{Stage I: Plan}, \textbf{Stage II: Query}, \textbf{Stage III: Request}, and \textbf{Stage IV: Reply} (\autoref{fig:MASPipeLine}). Detailed prompts for each stage are provided in \autoref{app:prompt}.

\par \textbf{Stage I: Plan.} The goal of this stage is to develop a list of strategy objectives, referred to as the \textit{Plan}. Each agent begins by analyzing its existing \textit{SC} structure and integrating both global and user-defined \textit{Knowledge}. Based on this analysis, it formulates a set of \textit{Plans}, each accompanied by explicit \textit{Reasons} to ensure traceability and transparency in decision intent.

\par \textbf{Stage II: Query.} This stage identifies potential partners, or \textit{Candidates}, for each \textit{Plan}. Agents translate their \textit{Plans} into specific partnership requirements, expressed as \textit{Constraints}. They define industry boundaries, desired attributes, and corresponding weights. Using predefined query functions, each agent retrieves and ranks a list of companies that best satisfy the weighted criteria, forming the initial \textit{Candidate} pool.

\par \textbf{Stage III: Request.} In this stage, agents formalize partner decisions by generating concrete \textit{Requests}. For expansion-oriented \textit{Plan}, agents evaluate their \textit{Candidates} and select suitable companies as collaboration targets. For each target, they craft persuasive, context-aware \textit{Request} that reference mutual benefits and alignment with the partner's profile. Conversely, if a \textit{Plan} involves reducing collaborations (e.g., ``removing suppliers with low innovation''), agents issue termination \textit{Requests} accordingly.

\par \textbf{Stage IV: Reply.} The final stage focuses on processing incoming \textit{Requests} from other agents. Each agent assesses the received proposals by comparing them with its own \textit{SC} and active \textit{Plans}. The evaluation considers strategic alignment, feasibility, and associated risks. Based on this assessment, the agent either accepts or declines each \textit{Request}, generating a corresponding \textit{Responses} with articulated \textit{Reasons}. This mechanism fosters transparency, rationality, and accountability in partnership formation.

\subsubsection{Supply Chain Explain Model and XAI}
\par Building on the above simulation results, we developed a predictive modeling module to examine how structural dynamics within the SC influence firm-level performance outcomes (\textbf{DR4}). Given the heterogeneity of strategic behaviors across firms, we trained individualized models using temporal SC data (\textbf{DR3}). The input features include node attributes and first-order network connections, while the prediction targets correspond to firm-level performance metrics. Network embeddings are generated using a Variational Graph Autoencoder, a classical approach for unsupervised learning on graph-structured data~\cite{kipf2016variationalgraphautoencoders}. To accommodate variations across datasets, users can flexibly select among multiple regression models, including \textit{Linear Regression}, \textit{Lasso Regression}, \textit{Random Forests (RF)} and \textit{GradientBoosting}, based on performance metrics and analytical preferences.

\par To ensure interpretability (\textbf{DR4}), we incorporated SHAP to quantify the contributions of individual features. SHAP analysis offers two complementary perspectives: at the global level, it identifies key collaborators and structural attributes that most strongly influence firm performance; at the local level, it elucidates how specific factors affect predictions for individual companies. This quantitative layer complements the qualitative CoT reasoning, jointly providing a transparent and comprehensive understanding of agent-driven decision outcomes.

\subsubsection{Horizon Extend Model}
\par To enable simulations beyond the temporal scope of existing data (\textbf{DR2}), \textit{SCSimulator} incorporates a forecasting component termed the \textit{Horizon Extend Model}. Its primary goal is to extrapolate enterprise-level features that evolve over time yet remain independent of network structure. Starting from the final timestamp of the provided dataset, the model predicts feature values at the next time step $t_{n+1}$ based on the observed series $t_1, \cdots, t_n$. To ensure adaptability across different datasets, users can flexibly choose among several regression models, including \textit{Linear Regression}, \textit{Lasso Regression}, \textit{RF}, and \textit{Gradient Boosting}, based on performance outcomes and analytical needs.


\begin{figure*}[h]
    \centering
    \includegraphics[width=\linewidth]{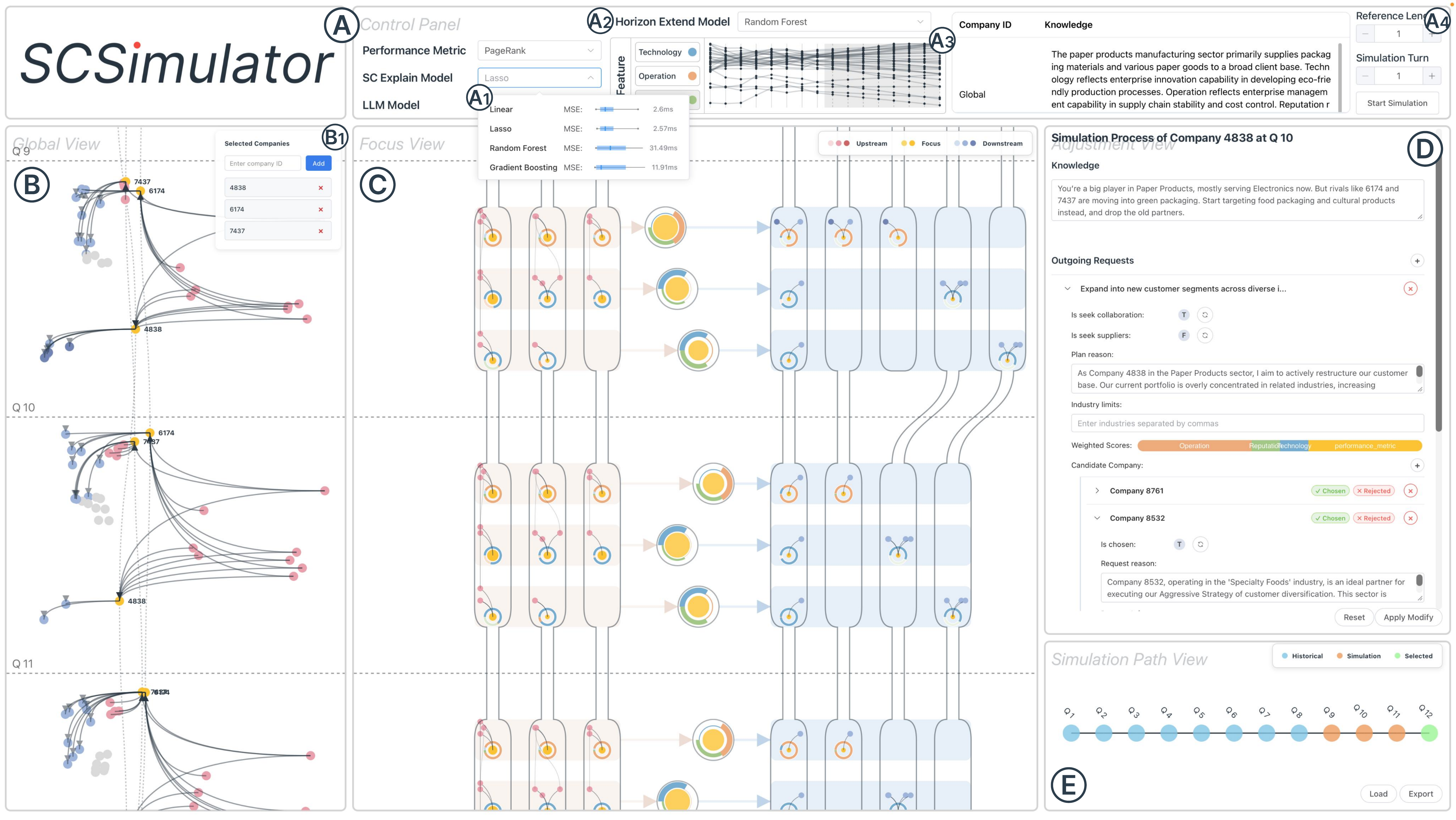}
    \caption{The \textit{SCSimulator} simulation interface incorporates five main views, each serving a distinct purpose. \raisebox{-0.08cm}{\includegraphics[height=0.31cm]{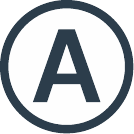}} The \textit{Control Panel} aims to support model selection and parameter configuration. \raisebox{-0.08cm}{\includegraphics[height=0.31cm]{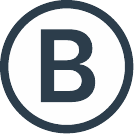}} The \textit{Global View} aims to reveal the temporal evolution and relational dynamics of the SC structure. \raisebox{-0.08cm}{\includegraphics[height=0.31cm]{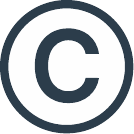}} The \textit{Focus View} aims to reveal how a focal enterprise interacts with its partners, highlighting influence balance and relationship dynamics within SCs. \raisebox{-0.08cm}{\includegraphics[height=0.31cm]{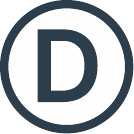}} The \textit{Adjustment View} aims to trace and adjust agents’ behaviors and interactions throughout the simulation. \raisebox{-0.08cm}{\includegraphics[height=0.31cm]{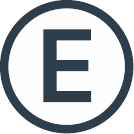}} The \textit{Simulation Path View} aims to visualize simulation progress and enable exploration of alternative paths.}
    \label{fig:system}
\end{figure*}

\subsection{Frontend Interface}

\subsubsection{Upload Page}

\par
\begin{figure}[h]
    \centering
    \includegraphics[width=\linewidth]{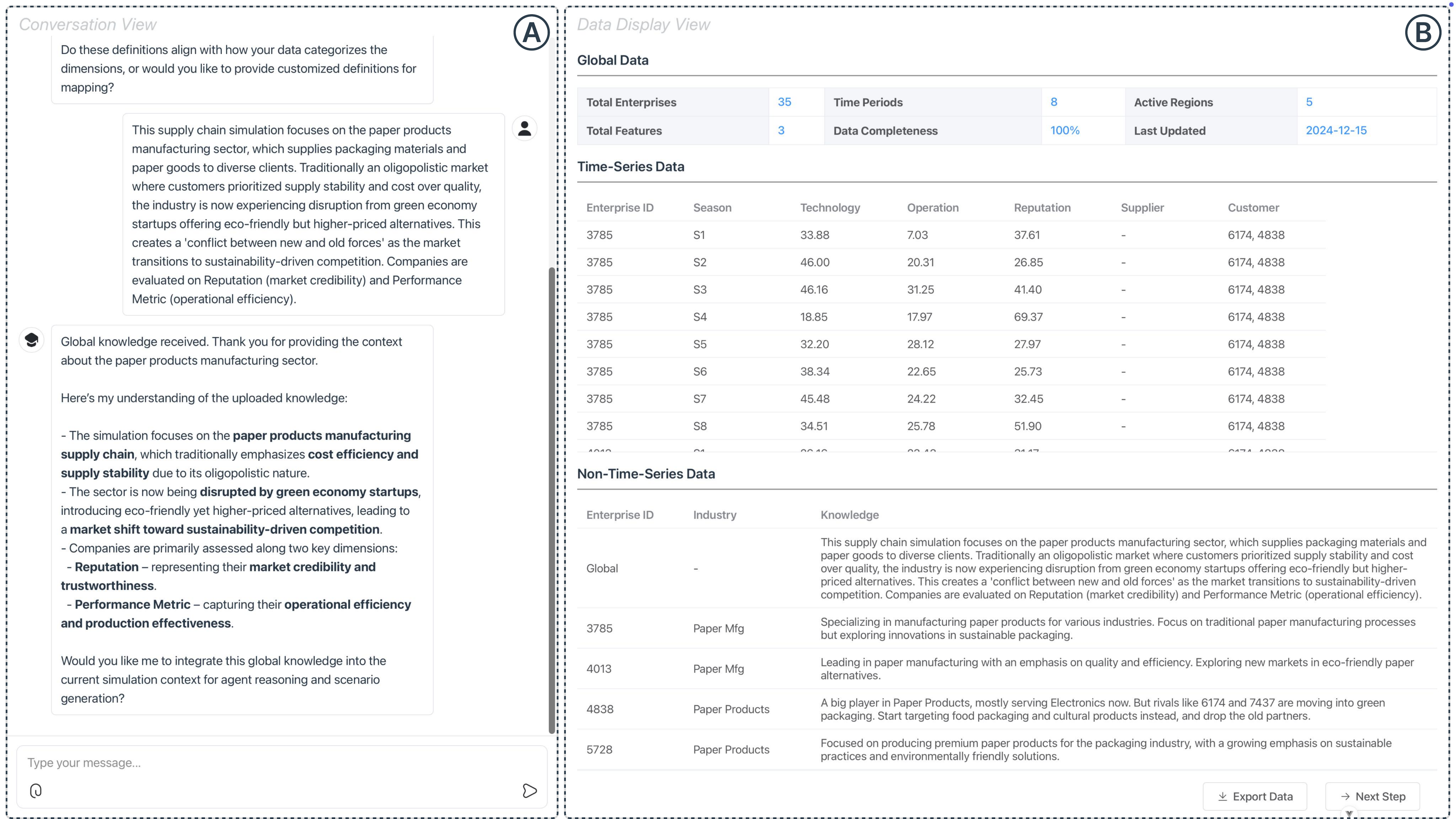}
    \caption{The \textit{Upload Page} of \textit{SCSimulator}. Users upload raw datasets and collaborate with the LLM agent through the \textit{Conversation View} \raisebox{-0.08cm}{\includegraphics[height=0.31cm]{figure/A.pdf}} to map data into structured templates. The \textit{Data Display View} \raisebox{-0.08cm}{\includegraphics[height=0.31cm]{figure/B.pdf}} presents real-time mapping results for review and modification before proceeding to the next step.}
    \label{fig:uploadPage}
\end{figure}

 On the \textit{Upload Page} (\autoref{fig:uploadPage}), the user can upload their original datasets and interact with the agent through the \textit{Conversation View} (\autoref{fig:uploadPage}-\raisebox{-0.09cm}{\includegraphics[height=0.35cm]{figure/A.pdf}}) to map raw data into the \textit{SCSimulator} template (\textbf{DR1}). The mapping results are displayed in real time within the \textit{Data Display View} (\autoref{fig:uploadPage}-\raisebox{-0.09cm}{\includegraphics[height=0.35cm]{figure/B.pdf}}), where users can directly review and modify the mappings. Upon completion, users can either click \raisebox{-0.09cm}{\includegraphics[height=0.35cm]{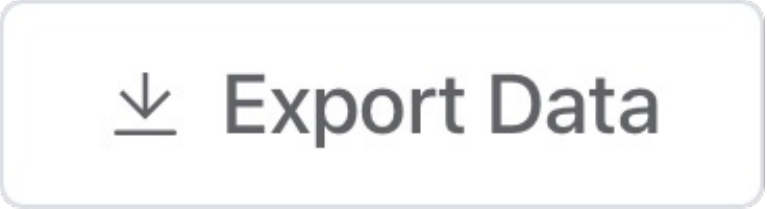}} (Export Data) to return to the main system page or \raisebox{-0.09cm}{\includegraphics[height=0.35cm]{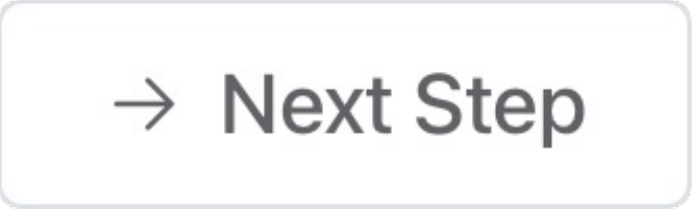}} (Next Step) to proceed with exporting the results.

\subsubsection{Control Panel}
\par The \textit{Control Panel} provides users with the ability to configure key simulation parameters, ensuring controlled and flexible execution (\textbf{DR1}). Users can specify the Performance Metric, SC Explain Model, Horizon Extend Model, and Agent Model that drive agent behavior. For the SC Explain Model (\autoref{fig:system}-\raisebox{-0.09cm}{\includegraphics[height=0.35cm]{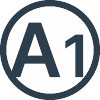}}) and Horizon Extend Model (\autoref{fig:system}-\raisebox{-0.09cm}{\includegraphics[height=0.35cm]{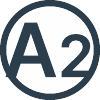}}), box plots visualize model performance and runtime, assisting users in selecting models that best balance accuracy and computational cost.

\par In the Horizon Extend Model view (\autoref{fig:system}-\raisebox{-0.09cm}{\includegraphics[height=0.35cm]{figure/A2.pdf}}), an interactive visualization supports exploration of the extended feature space. On the left, the Feature Bar allows users to toggle between enterprise features. The line chart illustrates the temporal evolution of selected company features, with white regions representing historical data and gray regions denoting predictions generated by the Horizon Extend Model.

\par To the right, the Knowledge table (\autoref{fig:system}-\raisebox{-0.09cm}{\includegraphics[height=0.35cm]{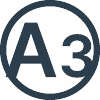}}) enables users to inspect and directly edit underlying Knowledge representations. The rightmost section (\autoref{fig:system}-\raisebox{-0.09cm}{\includegraphics[height=0.35cm]{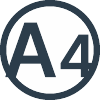}}) allows users to configure the Reference Length, i.e., the number of previous time steps agents consider, and the Simulation Turns, indicating how many iterations the simulation will execute. By clicking \raisebox{-0.09cm}{\includegraphics[height=0.35cm]{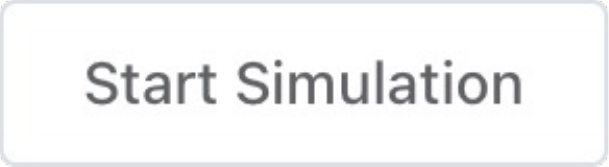}} (Start Simulation), users can initiate the backend process. This design streamlines configuration and execution, facilitating efficient and transparent simulation control.

\begin{figure*}[h]
    \centering
    \includegraphics[width=\linewidth]{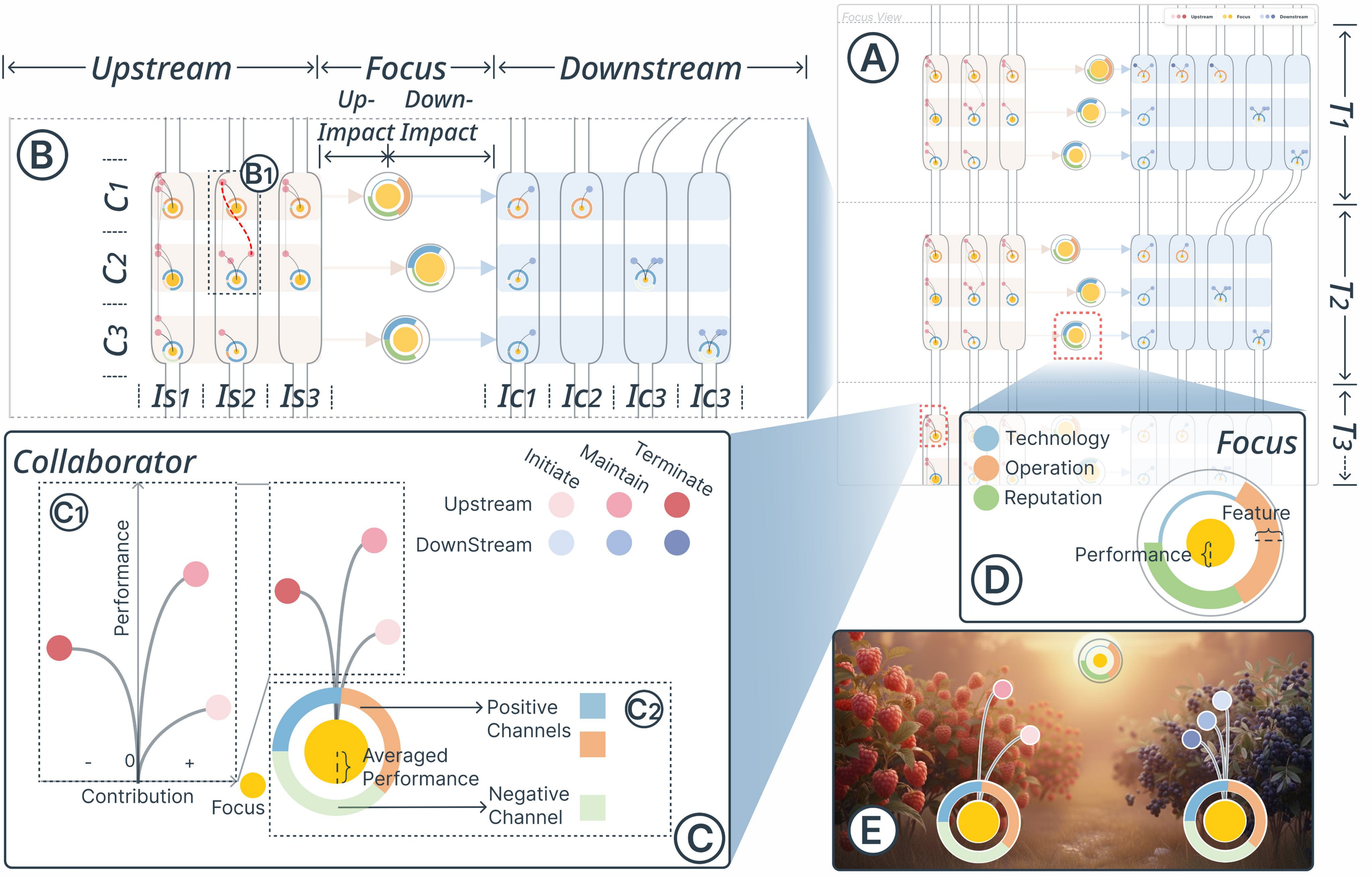}
    \caption{The \textit{Focus View} Design. \raisebox{-0.08cm}{\includegraphics[height=0.31cm]{figure/A.pdf}} The overall layout arranges elements chronologically from top to bottom. \raisebox{-0.08cm}{\includegraphics[height=0.31cm]{figure/B.pdf}} At each timestamp, suppliers are on the left, the focal company in the center, and customers on the right; each row shows one focal company's SC for comparison. Red dashed lines \raisebox{-0.08cm}{\includegraphics[height=0.31cm]{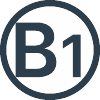}} mark shared suppliers across focal companies. \raisebox{-0.08cm}{\includegraphics[height=0.31cm]{figure/C.pdf}} Industry-grouped partners use a berry metaphor: raspberries (warm tones) for upstream suppliers and blackberries (cool tones) for downstream customers; the lower "soil" half shows the focal company's influence on the group, while the upper "berries" half depicts individual partners' influence on the focal company. Colors encode collaboration stages (initiate, maintain, terminate). \raisebox{-0.08cm}{\includegraphics[height=0.31cm]{figure/D.pdf}} The focal company uses a rose chart glyph: the central yellow circle encodes performance (larger radius for higher values), and outer arcs represent features like technology, operation, and reputation (radii encode magnitudes). \raisebox{-0.08cm}{\includegraphics[height=0.31cm]{figure/E.pdf}} The metaphor overview: the focal company as the sun, with upstream and downstream partners as berry bushes on either side.}
    \label{fig:focusView}    
\end{figure*}

\subsubsection{Global View}
\par The \textit{Global View} (\autoref{fig:system}-\raisebox{-0.09cm}{\includegraphics[height=0.35cm]{figure/B.pdf}}) provides a top-down overview of the temporal evolution of the SC structure (\textbf{DR5}). At each timestamp, every company is projected onto a two-dimensional plane based on both its intrinsic features and those of its partners. To preserve spatial consistency across time, data from all timestamps are concatenated and jointly subjected to dimensionality reduction. The resulting embeddings are then displayed according to their respective temporal intervals.

\par Users can hover over a node to highlight the selected company along with its upstream and downstream partners. These firms are connected through counterclockwise arcs from suppliers to customers to minimize visual overlap. The same company appearing across different timestamps is linked by a dashed line. The hovered company is shown in light yellow, while its upstream and downstream partners are colored with warm and cool tones, respectively. Varying color saturation indicates the lifecycle of partnership, from initiation, through continuation, to termination, following a garden metaphor, which will be elaborated in (\autoref{sec:sys_fv}). Users can click a node to add it to the selected company list, which highlights it in deep yellow. The selected companies are shown in a dedicated list (\autoref{fig:system}-\raisebox{-0.09cm}{\includegraphics[height=0.35cm]{figure/B1.pdf}}), where users can add or remove entries by searching their company IDs.
\label{sec:sys_fv}

\subsubsection{Focus View}

\par The \textit{Focus View} (\autoref{fig:system}-\raisebox{-0.09cm}{\includegraphics[height=0.35cm]{figure/C.pdf}}) is designed to enable users to observe network dynamics from the perspective of a specific enterprise (\textbf{DR4,5}). Like the \textit{Global View}, it arranges elements chronologically (\autoref{fig:focusView}-\raisebox{-0.09cm}{\includegraphics[height=0.35cm]{figure/A.pdf}}). At each timestamp (\autoref{fig:focusView}-\raisebox{-0.09cm}{\includegraphics[height=0.35cm]{figure/B.pdf}}), enterprises are arranged vertically ($C_1$ - $C_3$), with suppliers and upstream entities positioned on the left, the focal enterprise at the center, and customers and downstream entities on the right. Warm and cool color schemes differentiate the upstream and downstream regions. Within each region, companies are grouped by industry ($I_{s1}$ - $I_{s3}$ for suppliers and $I_{c1}$ - $I_{c4}$ for customers). The horizontal position of the central focus node (\autoref{fig:focusView}-\raisebox{-0.09cm}{\includegraphics[height=0.35cm]{figure/D.pdf}}), representing the focal enterprise, encodes the relative influence of upstream and downstream actors, computed from SHAP values as the ratio between the summed absolute SHAP contributions of suppliers and customers. Nodes closer to the left indicate stronger supplier influence, centered positions indicate balance, and nodes closer to the right indicate stronger customer influence.

\par To illustrate the interactions between the focal company and its upstream and downstream partners, we introduce a ``berry orchard'' metaphor (\autoref{fig:focusView}-\raisebox{-0.09cm}{\includegraphics[height=0.35cm]{figure/E.pdf}}). In this metaphor, the focal company is positioned at the center, represented as the sun, while its partners are symbolized as berries on either sides. Raspberries, in warm tones, denote upstream suppliers, and blackberries, in cool tones, represent downstream customers.

\par For the focal company, a rose chart–based ring glyph (\autoref{fig:focusView}-\raisebox{-0.09cm}{\includegraphics[height=0.35cm]{figure/D.pdf}}) highlights key attributes. The central yellow circle encodes the performance metric, with a larger radius indicating higher performance. The outer arcs represent various company features, such as technology, operation, and reputation. These features are color-coded to differentiate their types, while the radii encode their magnitudes.

\par Each berry glyph aggregates firms within an industry sector. The lower half, referred to as ``soil'', represents the focal company's influence on the group  (\autoref{fig:focusView}-\raisebox{-0.09cm}{\includegraphics[height=0.35cm]{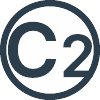}}) , while the upper half, ``berries'', represents specific partner firms, showing their influence on the focal company  (\autoref{fig:focusView}-\raisebox{-0.09cm}{\includegraphics[height=0.35cm]{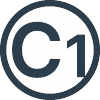}}). Within the soil, the central yellow circle encodes the group's average performance, while a surrounding ring summarizes the focal company's contribution. The opacity and angular span of this ring indicate the polarity and relative magnitude of these effects, computed from the aggregated SHAP values of the corresponding attributes.

\par Each berry corresponds to an individual company. The vertical position of each berry reflects performance, while its horizontal position shows its contribution to the focal company, determined by its SHAP value. Closer positions indicate a positive influence, while farther positions suggest a negative influence. For suppliers, berries positioned farther to the right imply greater benefit to the focal company, while for customers, berries farther to the left imply the same. The color gradient of each berry encodes the collaboration stages, ranging from light (initiate) to dark (terminate), ensuring consistency between raspberries in warm tones and blackberries in cool tones.

\par In our design, a robust SC is characterized by balanced berry clusters on both sides, with multiple fruits, signifying structural resilience and mutual growth. Darker soil rings indicate that the focal company actively contributes to the development of its partners, fostering reciprocal relationships. Additionally, red dashed lines between supplier clusters (\autoref{fig:focusView}-\raisebox{-0.09cm}{\includegraphics[height=0.35cm]{figure/B1.pdf}}) highlight shared suppliers between firms, such as $C_1$ and $C_2$.

\begin{figure}[t]
  \centering
  \includegraphics[width=0.8\linewidth]{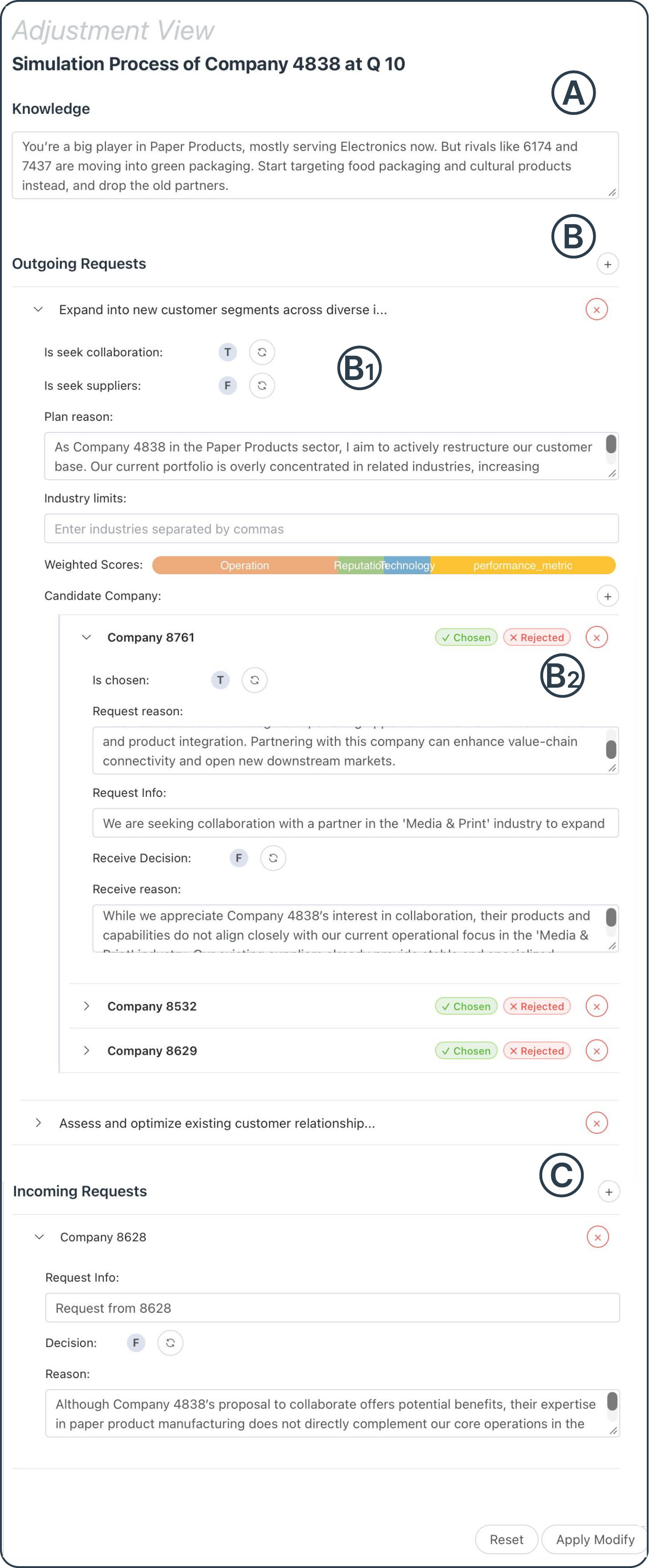}
    \caption{The \textit{Adjustment View} design. \raisebox{-0.08cm}{\includegraphics[height=0.31cm]{figure/A.pdf}} Knowledge panel displays the agent's current Knowledge. \raisebox{-0.08cm}{\includegraphics[height=0.31cm]{figure/B.pdf}} Outgoing Requests section lists proactive Plans in a collapsible format; expanded views show Plan type, Reasons, and targets \raisebox{-0.08cm}{\includegraphics[height=0.31cm]{figure/B1.pdf}}, with Candidate details including company ID, selection status, response outcome, and dialogues \raisebox{-0.08cm}{\includegraphics[height=0.31cm]{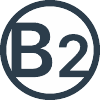}}. \raisebox{-0.08cm}{\includegraphics[height=0.31cm]{figure/C.pdf}} Incoming Requests section mirrors the structure for requests received from other companies, enabling review of details and decisions.}
  \label{fig:adjustmentView}
\end{figure}

\subsubsection{Adjustment View}

\par The \textit{Adjustment View} (\autoref{fig:system}-\raisebox{-0.09cm}{\includegraphics[height=0.35cm]{figure/D.pdf}}) enables users to modify specific company agents at selected timestamps (\textbf{DR6}). To minimize cognitive load from the extensive simulation data, we have designed a tree-like structure for both visualization and interaction. The interface comprises three main components: the Knowledge panel (\autoref{fig:adjustmentView}-\raisebox{-0.09cm}{\includegraphics[height=0.35cm]{figure/A.pdf}}), the Outgoing Requests section (\autoref{fig:adjustmentView}-\raisebox{-0.09cm}{\includegraphics[height=0.35cm]{figure/B.pdf}}), and the Incoming Requests section (\autoref{fig:adjustmentView}-\raisebox{-0.09cm}{\includegraphics[height=0.35cm]{figure/C.pdf}}). In the Knowledge panel, users can review the agent's current Knowledge as needed. The Outgoing Requests section displays the agent's proactive actions in a Plan list, which summarizes all Plans in a collapsed format. Users can expand each plan to view its type, rationale, targets, and a list of Candidates (\autoref{fig:adjustmentView}-\raisebox{-0.09cm}{\includegraphics[height=0.35cm]{figure/B1.pdf}}). Each Candidate entry provides the company ID, selection status, and response outcome. Users can expand these entries further to explore detailed dialogues (\autoref{fig:adjustmentView}-\raisebox{-0.09cm}{\includegraphics[height=0.35cm]{figure/B2.pdf}}). Similarly, the Incoming Requests section follows the same structure, showing companies that have initiated requests to the current agent and their corresponding Replys. Users can intervene in both Plans and Candidates using dedicated controls: \raisebox{-0.09cm}{\includegraphics[height=0.35cm]{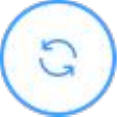}} (Negate) to refresh rejected collaborations, \raisebox{-0.09cm}{\includegraphics[height=0.35cm]{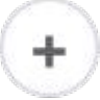}} (Add) to add or \raisebox{-0.09cm}{\includegraphics[height=0.35cm]{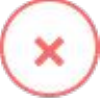}} (Delete) to remove. If a modification is unsatisfactory, users can click \raisebox{-0.09cm}{\includegraphics[height=0.35cm]{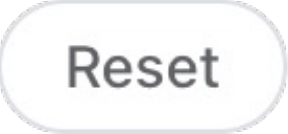}} (Reset) to revert to the initial state, or \raisebox{-0.09cm}{\includegraphics[height=0.35cm]{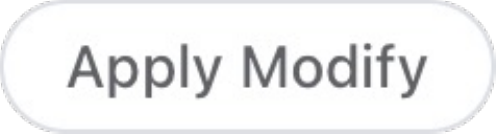}} (Apply Modify) to confirm the changes. All confirmed changes are recorded with new timestamps and reflected in the \textit{Simulation Path} view, allowing users to trace the simulation's evolution over time.

\subsubsection{Simulator Path View}

\par The \textit{Simulation Path View} is designed to help users track the progress of their simulation process (\textbf{DR7}). A tree diagram is used to visualize the simulation paths, with each node representing a simulation turn. Nodes are color-coded to indicate different statuses: \raisebox{-0.09cm}{\includegraphics[height=0.35cm]{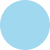}} for historical data, \raisebox{-0.09cm}{\includegraphics[height=0.35cm]{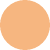}} for simulated nodes, and \raisebox{-0.09cm}{\includegraphics[height=0.35cm]{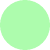}} for the currently active node. The x-coordinate reflects the temporal sequence, with nodes connected from left to right to illustrate progression. Users can click on nodes to switch between paths. Once satisfied, they can click \raisebox{-0.09cm}{\includegraphics[height=0.35cm]{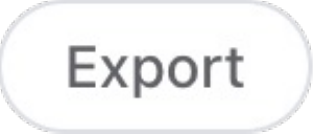}} (Export) to export and download the simulation log.

\section{Technical Evaluation}
\label{sec:technical_evaluation}

\par We conducted a technical evaluation of our LLM-driven agent simulation framework (\autoref{sec:agent-framework}) to assess its ability to model SC partner selection. We constructed a SC dataset in collaboration with domain experts from the formative study, based on real-world data from the upstream and downstream paper packaging industry. The dataset integrates SC network structure and firm-level attributes. Raw data were collected from public sources\footnote{\url{http://cebpubservice.cn/}; \url{https://www.gsxt.gov.cn/}}, including tendering and bidding relationships as network edges, and firm profiles as node attributes. Firm-level features include both non-temporal attributes, e.g., industry,  and temporal attributes, e.g., patents. Temporal attributes were categorized and scored by experts along three dimensions—\textit{Operation}, \textit{Technology}, and \textit{Reputation}, using a normalized 0–100 scale. The final dataset comprises 35 anonymized firms observed over eight quarters from 2023 to 2024.

\par To comprehensively assess the proposed framework, we evaluated simulation outcomes using a combination of quantitative performance metrics and expert-based subjective assessments.

\subsection{Quantitative Evaluation of Simulation Outcomes}
\par In each run, we simulated the decision-making of a single focal firm without user prompt intervention. We conducted $80$ independent runs on $4$ key nodes with fully observed data. $4$ consecutive time steps served as historical input, and the simulated network evolution was evaluated against the observed network at the subsequent step. Firms collaborating with the focal firm were labeled as $1$ and others as $0$, from which we report Accuracy (ACC), Precision, Recall, and F1-score.

\par To evaluate decision consistency, edge dynamics were categorized as \textit{add}, \textit{remove}, or \textit{keep}. Due to strong class imbalance, we adopted Gwet’s AC1 \cite{gwet2008computing} as a global agreement metric. At the firm level, we additionally report a consistency ratio (CR), defined as the proportion of the most frequent decision category. We categorized CR values into three levels: high CR (CR $>$ 0.8), medium CR (0.6 $<$ CR $\leq$ 0.8), and low CR (CR $\leq$ 0.6). Quantitative results are summarized in \autoref{table:quantitative_analysis}.

\par Overall, the simulation achieves moderate alignment with historical observations. This outcome is expected, as the framework is not intended for deterministic prediction but for exploring plausible decision trajectories within MAS. From this perspective, moderate predictive accuracy is an appropriate objective.

\par Agents exhibit high overall decision consistency, which can be attributed to the absence of prompt intervention and the imbalance in edge dynamics. A subset of firms shows medium-level consistency, potentially reflecting structurally influential or strategically pivotal actors whose decisions introduce greater variability.

\par Performance differences across the three LLMs are limited. While all models perform comparably, ChatGPT-4o shows a small but consistent advantage in performance and is therefore selected as the default backbone model for subsequent experiments. Overall, these results support the suitability of the proposed framework for simulating firm-level partner selection within SCs.

\begin{table*}[h]
\centering
\caption{Quantitative performance and decision consistency of the LLM-driven agent simulation across three LLMs, ChatGPT-4o (ChatGPT), Gemini-2.5-flash (Gemini) and Qwen3-VL-Flash (Qwen3). ACC, Precision, Recall, and F1-score measure similarity between simulated and observed network evolution, while Gwet’s AC1 and consistency ratios (CR) capture agreement in agent decision dynamics.}
\label{table:quantitative_analysis}
{\small
\begin{tabular}{lllllllll}
\hline
\textbf{Model} & \textbf{ACC (\%)} & \textbf{Precision (\%)} & \textbf{Recall (\%)} & \textbf{F1 (\%)} & \textbf{Gwet's AC1 (\%)} & \textbf{High CR (\%)} & \textbf{Medium CR (\%)} & \textbf{Low CR (\%)} \\ \hline
ChatGPT     & 72.43             & 68.20                   & 59.75                & 63.11            & 85.17                    & 91.18                 & 7.35                    & 1.47                 \\
Gemini         & 69.85             & 61.92                   & 58.83                & 59.47            & 81.72                    & 84.55                 & 14.71                   & 0.74                  \\
Qwen3        & 66.57             & 59.38                   & 48.56                & 52.67            & 98.21                    & 98.52                 & 0.74                    & 0.74             \\ \hline    
\end{tabular}
}
\end{table*}

\subsection{Expert-Based Subjective Evaluation}
\begin{table*}[h]
\centering
\caption{Expert-based evaluation of agent behavior across different simulation stages. Ratings were collected using a 5-point Likert scale, with higher scores indicating stronger agreement.}
\label{tab:expert-eval}
\begin{tabular}{lll}\hline
\textbf{Stage} & \textbf{Question}                                                             & \textbf{Mean / S.D.} \\\hline
Stage 1        & Q1: The agent can formulate reasonable plans.                                     & 4.3 / 0.67           \\
Stage 2 \& 3   & Q2 :The agent reasonably selected candidate companies as its targets.             & 3.7 / 0.67           \\
Stage 4        & Q3: The responses of other agents to its actions are reasonable.                  & 4.7 / 0.48           \\
All            & Q4: The agent closely adhered to the guidance provided by the external knowledge. & 4.4 / 0.83          \\ \hline
\end{tabular}
\end{table*}

\par To complement the quantitative results, we conducted an expert-based evaluation with ten master’s students trained in management, who acted as domain experts. Participants were allowed to configure external knowledge to intervene in the simulation and evaluated system behavior at different stages, as well as agents’ adherence to the provided knowledge, using 5-point Likert scales. Each expert independently conducted and reviewed five simulation rounds.

\par As summarized in \autoref{tab:expert-eval}, experts reported overall positive assessments of the system. Among the items, Q2 received comparatively lower ratings. Qualitative observations suggest two recurring behaviors: without expert knowledge, agents tended to act conservatively, formulating adjustment strategies but occasionally avoiding concrete partner selection; with expert knowledge, agents sometimes overreacted, issuing an excessive number of collaboration requests following strategic changes.

\par Overall, experts considered the framework credible and useful for exploring SC partner selection, particularly as a tool for sensemaking and scenario analysis.

\section{Usage Scenario}
\par In this section, we present a usage scenario to illustrate the workflow of \textit{SCSimulator} and demonstrate its effectiveness. The scenario is based on a case performed by \textbf{P1} and \textbf{P7} in our user study (see \autoref{sec:userStudy}), with minor modifications made for illustrative purposes.

\begin{figure*}[h]
  \centering
  \includegraphics[width=\linewidth]{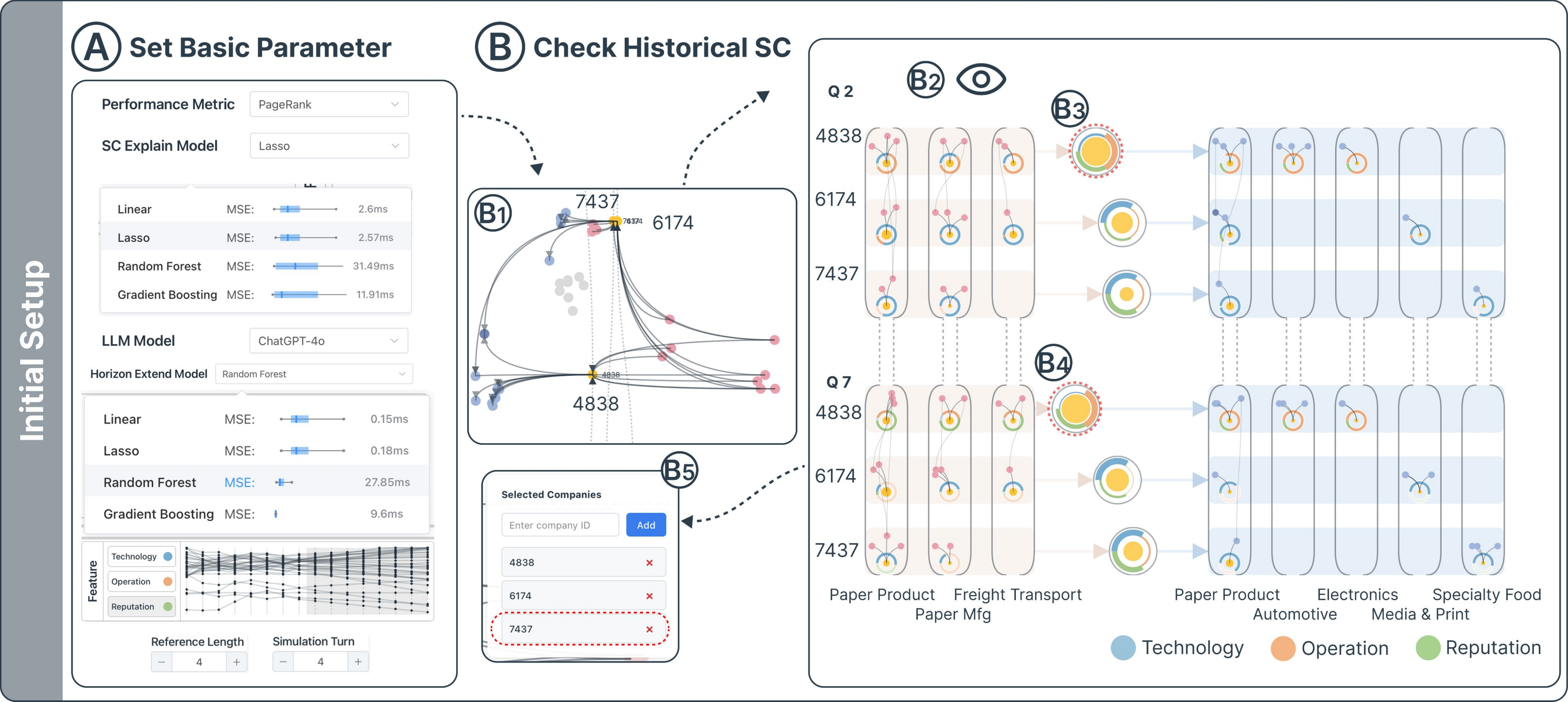}
    \caption{Initial Setup. \raisebox{-0.08cm}{\includegraphics[height=0.31cm]{figure/A.pdf}} Users set basic parameters, including performance metrics, LLM agents, and simulation settings. \raisebox{-0.08cm}{\includegraphics[height=0.31cm]{figure/B.pdf}} Users check historical SC data. \raisebox{-0.08cm}{\includegraphics[height=0.31cm]{figure/B1.pdf}} Users select companies of interest based on their SC network structure. \raisebox{-0.08cm}{\includegraphics[height=0.31cm]{figure/B2.pdf}} Users examine the supply and customer base of selected companies in the \textit{Focus View}. \raisebox{-0.08cm}{\includegraphics[height=0.31cm]{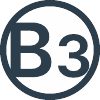}} $\&$ \raisebox{-0.08cm}{\includegraphics[height=0.31cm]{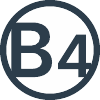}} Users analyze the competitive landscape and identify emerging competitors. \raisebox{-0.08cm}{\includegraphics[height=0.31cm]{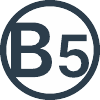}} Users confirm the list of selected companies for further analysis.}
  \label{fig:usage_scenario_1}
\end{figure*}

\subsection{Background}
\par The paper products sector traditionally emphasizes cost efficiency and supply stability, resulting in a relatively oligopolistic market structure. Recent growth in the green economy has introduced new entrants offering environmentally transparent products at premium prices, reshaping competitive dynamics and increasing uncertainty for incumbent firms~\cite{Maida_2024}. This shift has drawn expert attention to strategic tensions between established manufacturers and emerging competitors.

\subsection{Initial Setup}



\par Alice, a senior manager at \textbf{Company-4838} (\textbf{4838}), initiates her analysis using the LLM-based interface of \textit{SCSimulator}. After loading historical SC data, she configures the simulation by selecting \textbf{PageRank} as the performance metric, \textbf{ChatGPT-4o} as the LLM agent, \textbf{Lasso Regression} for the SC Explain Model, and \textbf{Random Forest} for horizon extension. She sets both the reference window and simulation turn length to four quarters.

\begin{figure*}[h]
  \centering
  \includegraphics[width=\linewidth]{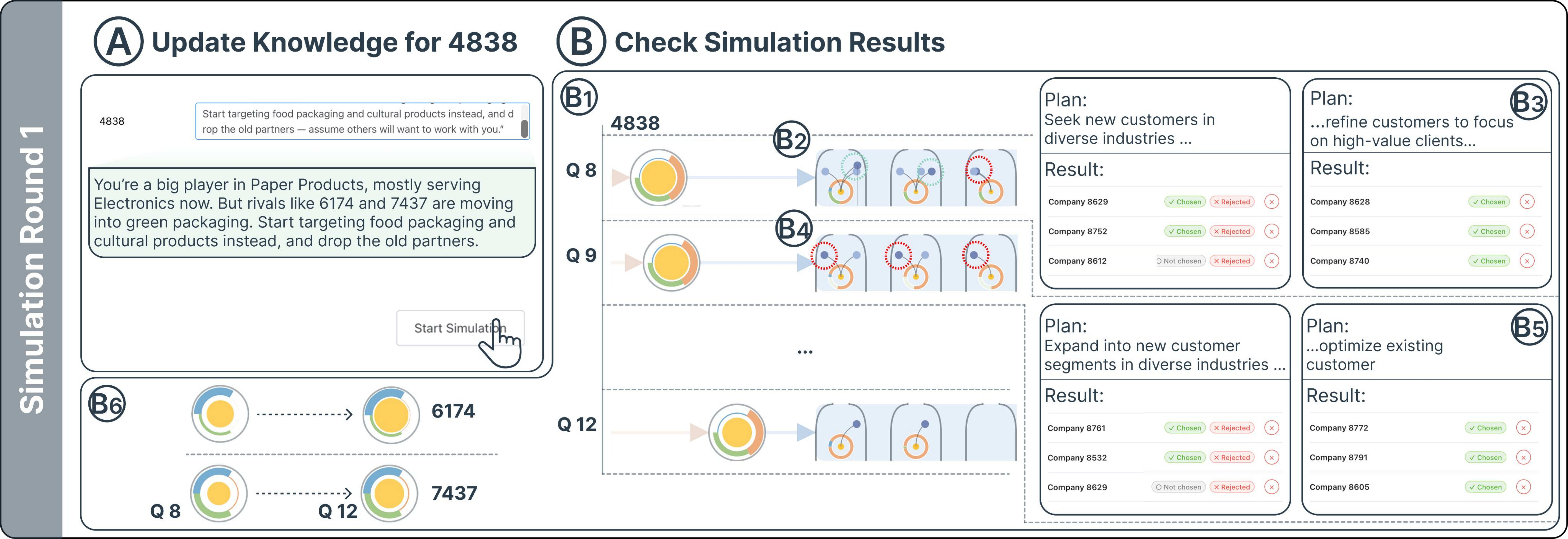}
\caption{Simulation Round 1: Radical Strategic Experimentation. 
\raisebox{-0.08cm}{\includegraphics[height=0.31cm]{figure/A.pdf}} Users update the Knowledge parameter for \textbf{4838} with a radical strategy to target new customers in green sectors and drop old partners. 
\raisebox{-0.08cm}{\includegraphics[height=0.31cm]{figure/B.pdf}} Users check simulation results. 
\raisebox{-0.08cm}{\includegraphics[height=0.31cm]{figure/B1.pdf}} \textbf{4838}'s overall performance declines significantly from Q8 to Q12. 
\raisebox{-0.08cm}{\includegraphics[height=0.31cm]{figure/B2.pdf}} \textbf{4838} initially conducts a reasonable refinement of its SC by removing poorly performing partners while preserving firms beneficial to its long-term development. 
\raisebox{-0.08cm}{\includegraphics[height=0.31cm]{figure/B3.pdf}} In Q9, the strategy of seeking new customers and optimizing existing ones fails to gain new partners.
\raisebox{-0.08cm}{\includegraphics[height=0.31cm]{figure/B4.pdf}} In Q9, excessive and misaligned refinement leads to the removal of developmentally valuable partners, triggering a sharp performance decline.
\raisebox{-0.08cm}{\includegraphics[height=0.31cm]{figure/B5.pdf}} In Q10, a continued focus on expansion and optimization is unsuccessful, leading to an overall performance drop.
\raisebox{-0.08cm}{\includegraphics[height=0.31cm]{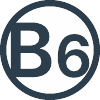}} Rivals \textbf{6174} and \textbf{7437} benefit from \textbf{4838}'s decline, gradually improving their relative positions in the network.}

  \label{fig:usage_scenario_2}
\end{figure*}


\par Exploring the \textit{Global View} (\autoref{fig:usage_scenario_1}), Alice selects \textbf{4838} and its established competitor \textbf{Company-6174} (\textbf{6174}). Network proximity analysis (\autoref{fig:usage_scenario_1}-\raisebox{-0.09cm}{\includegraphics[height=0.35cm]{figure/B1.pdf}}) reveals a third firm, \textbf{Company-7437} (\textbf{7437}), with a similar structural position to \textbf{6174}. Further inspection in the \textit{Focus View} (\autoref{fig:usage_scenario_1}-\raisebox{-0.09cm}{\includegraphics[height=0.35cm]{figure/B2.pdf}}) confirme that all three companies share similar upstream suppliers. However, \textbf{4838} primarily serves traditional manufacturing sectors such as electronics, while \textbf{6174} and \textbf{7437} focus on consumer-oriented green packaging industries, including food and media products.

\par Alice further notes that \textbf{4838} is positioned further left in the network, indicating a stronger dependence on upstream suppliers. Although \textbf{4838} retains a relative advantage, its SHAP-informed positioning (\autoref{fig:usage_scenario_1}-\raisebox{-0.09cm}{\includegraphics[height=0.35cm]{figure/B3.pdf}} and \autoref{fig:usage_scenario_1}-\raisebox{-0.09cm}{\includegraphics[height=0.35cm]{figure/B4.pdf}}) reveals increasing upstream influence and a gradual erosion of strategic independence compared to \textbf{6174} and \textbf{7437}. Consequently, Alice infers that \textbf{4838} should diversify its customer base toward greener, consumer-oriented markets and includes \textbf{7437} for deeper analysis (\autoref{fig:usage_scenario_1}-\raisebox{-0.09cm}{\includegraphics[height=0.35cm]{figure/B5.pdf}}).

\subsection{Simulation Round 1: Radical Strategic Experimentation}
\begin{figure*}[h]
  \centering
  \includegraphics[width=\linewidth]{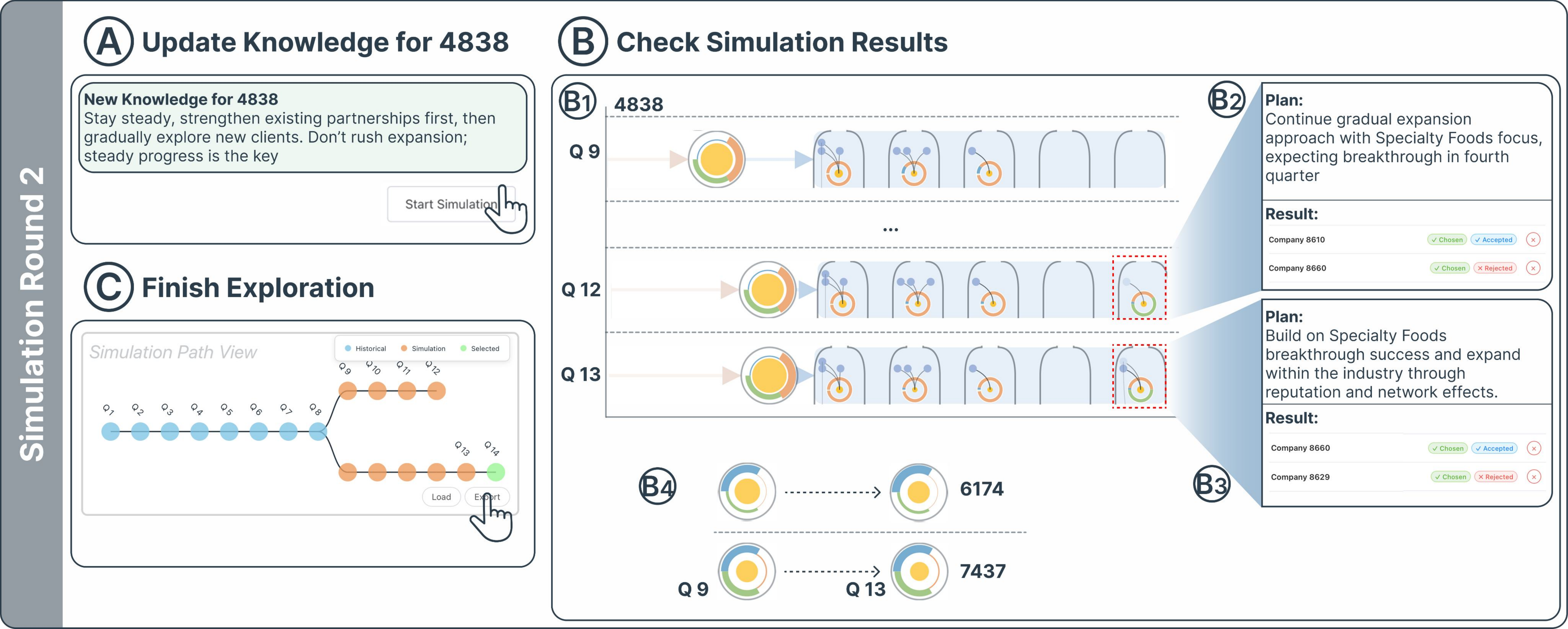}
    \caption{Simulation Round 2: Progressive Strategic Optimization. \raisebox{-0.08cm}{\includegraphics[height=0.31cm]{figure/A.pdf}} User updates \textbf{4838}'s Knowledge with a conservative strategy: strengthen existing partnerships first and then gradually explore new clients without rushing expansion. \raisebox{-0.08cm}{\includegraphics[height=0.31cm]{figure/B.pdf}} Simulation results show a positive turning point in \textbf{4838}'s customer acquisition. \raisebox{-0.08cm}{\includegraphics[height=0.31cm]{figure/B2.pdf}} 4838 secures its first new partnership with \textbf{8610} in Q12 after three quarters of effort, marking entry into the specialty food sector. \raisebox{-0.08cm}{\includegraphics[height=0.31cm]{figure/B3.pdf}} Building on this breakthrough, Q13 yield further partnerships, \textbf{8660}, in the same sector, demonstrating a network effect. \raisebox{-0.08cm}{\includegraphics[height=0.31cm]{figure/B4.pdf}} Rivals \textbf{6174} and \textbf{7437} remain inactive, allowing 4838 to solidify its position. \raisebox{-0.08cm}{\includegraphics[height=0.31cm]{figure/C.pdf}} The simulation path, including the initial radical experiment (Q8-Q12) and the subsequent progressive optimization (Q9-Q13), is recorded in the \textit{Simulation Path View}.}
  \label{fig:usage_scenario_3}
\end{figure*}

\par Alice recognized the growing threat posed by new competitors but remained confident in the market position of \textbf{4838}, particularly due to its strong performance in \textbf{Operation} (\autoref{fig:usage_scenario_2}-\raisebox{-0.09cm}{\includegraphics[height=0.35cm]{figure/B1.pdf}}). She began the first simulation round with a bold assumption: ``\textit{No company would refuse a strong partner.}''

\par To test this idea, Alice updated \textbf{4838}'s \textit{Knowledge} (\autoref{fig:usage_scenario_2}-\raisebox{-0.09cm}{\includegraphics[height=0.35cm]{figure/A.pdf}}). She instructed the agent to proactively target new customers in the specialty food and cultural product sectors, while discontinuing relationships with long-term but outdated partners. After configuring the parameters, she initiated the simulation.

\par In the early phase of the simulation, \textbf{4838} followed this strategy and eliminated several underperforming partners (highlighted by green circles), a decision that aligned with both its operational optimization goals and negative SHAP values (\autoref{fig:usage_scenario_2}-\raisebox{-0.09cm}{\includegraphics[height=0.35cm]{figure/B2.pdf}}). This initial pruning constituted a reasonable cleanup of low-value relationships and reflected a coherent alignment between data-driven signals and strategic intent.

\par However, \textbf{4838} failed to attract any new customers and instead experienced a noticeable decline in overall performance. A deeper investigation in the \textit{Adjustment View} (\autoref{fig:usage_scenario_2}-\raisebox{-0.09cm}{\includegraphics[height=0.35cm]{figure/B3.pdf}} and \autoref{fig:usage_scenario_2}-\raisebox{-0.09cm}{\includegraphics[height=0.35cm]{figure/B5.pdf}}) revealed that during Q9–Q10, 4838 focused
primarily on aggressive expansion and partial customer optimization.

\par Specifically, while the revised \textit{Knowledge} emphasized aggressive customer refinement and sectoral repositioning, it caused \textbf{4838} to terminate relationships with several companies that were beneficial to its long-term development. For instance, When dismissing \textbf{8605}, the agent explicitly noted that ``\textit{the low operation score conflicts with our strategic shift.}''. Although these firms exhibited relatively low \textit{Operation} scores, their SHAP values indicated positive contributions to \textbf{4838}'s performance trajectory (\autoref{fig:usage_scenario_2}-\raisebox{-0.09cm}{\includegraphics[height=0.35cm]{figure/B4.pdf}}). By strictly adhering to strategic guidance, the agent overlooked these latent benefits and removed partners that had empirically supported \textbf{4838}'s growth. Meanwhile, \textbf{4838} failed to obtain new customers due to its comparatively low technology score. For instance, \textbf{8752} declined cooperation, ``\textit{...\textbf{4838} has a significantly lower technology score compared to our current suppliers, which introduces unacceptable risks...}''. Within only two quarters, \textbf{4838} lost a substantial portion of its customer base without forming any new partnerships.

\par This decline further reinforced negative perceptions among other agents, who subsequently refused collaboration, commenting that ``\textit{...\textbf{4838}'s performance is continuing to decrease, indicating it is not a qualified supplier...}'' As a result, \textbf{4838} entered a downward spiral in which deteriorating performance discouraged new opportunities and amplified existing weaknesses. During the same period, competitors such as \textbf{6174} and \textbf{7437}, though largely inactive, benefited from \textbf{4838}'s missteps and gradually improved their relative positions within the network (\autoref{fig:usage_scenario_2}-\raisebox{-0.09cm}{\includegraphics[height=0.35cm]{figure/B6.pdf}}).

\par Upon reviewing the outcomes, Alice realized that while \textbf{4838} retained strong operational capacity, its decision-making process had become misaligned with the evolving industrial ecosystem. In particular, the excessive reliance on \textit{Knowledge}, without sufficient reconciliation with SHAP-based attribution signals, led to structurally harmful decisions during Q9. Rather than a failure of expansion alone, this round illustrates how the imbalance between normative strategic guidance and explainable, outcome-oriented evidence can destabilize otherwise competitive firms.

\subsection{Simulation Round 2: Progressive Strategic Optimization}

\par After reviewing the first-round outcomes, Alice realized that an overly aggressive knowledge input had caused \textbf{4838} to misjudge collaboration prospects and overestimate other firms’ willingness to accept its terms. In the second round, she adopted a more conservative strategy, discouraging reckless expansion and launching a new simulation (\autoref{fig:usage_scenario_3}-\raisebox{-0.09cm}{\includegraphics[height=0.35cm]{figure/A.pdf}}).

\par During the first three rounds, \textbf{4838}’s outreach attempts failed to elicit positive responses. A turning point emerged in Q12 (\autoref{fig:usage_scenario_3}-\raisebox{-0.09cm}{\includegraphics[height=0.35cm]{figure/B1.pdf}}), when \textbf{Company-8610} (\textbf{8610}), motivated by its own expansion needs, agreed to collaborate: ``\textit{...Due to our business expansion, we are willing to try collaboration with \textbf{4838}...}'' (\autoref{fig:usage_scenario_3}-\raisebox{-0.09cm}{\includegraphics[height=0.35cm]{figure/B2.pdf}}). This partnership marked \textbf{4838}’s entry into the specialty food industry and established an initial foothold. Building on this success, a subsequent round (Q13) led to a second collaboration with \textbf{Company-8660} (\textbf{8660}), which explicitly referenced \textbf{4838}’s prior cooperation with \textbf{8610} as a signal of credibility (\autoref{fig:usage_scenario_3}-\raisebox{-0.09cm}{\includegraphics[height=0.35cm]{figure/B3.pdf}}). These partnerships enabled efficient customer expansion, particularly as competitors \textbf{6174} and \textbf{7437} remained inactive (\autoref{fig:usage_scenario_3}-\raisebox{-0.09cm}{\includegraphics[height=0.35cm]{figure/B4.pdf}}).

\par The simulation indicates that entering a new market typically requires substantial upfront effort to secure an initial client, while subsequent partnerships incur lower marginal costs. This dynamic is consistent with real options theory, which interprets early investments as creating future growth opportunities~\cite{james2003option}. In contrast, NPV-based models prioritize short-term, predictable returns and often undervalue strategic flexibility under uncertainty~\cite{dixit1994investment,mcdonald2006role}, despite the potential for asymmetric upside.
\par Across two simulation rounds, Alice observed that although responsiveness to emerging trends is important, overly radical transformation can destabilize traditional firms. A gradual and conservative strategy better balances innovation and risk, echoing prior findings on the dual role of innovation~\cite{Wamba2023} and the effectiveness of conservative sales strategies in early market exploration~\cite{BorghSchepers2017}.

\par Finally, Alice exported the complete simulation log via the \textit{Simulation Path View} for further analysis (\autoref{fig:usage_scenario_3}-\raisebox{-0.09cm}{\includegraphics[height=0.35cm]{figure/C.pdf}}).

\section{User Study}
\label{sec:userStudy}

\par We conducted a user study with 12 participants to evaluate the effectiveness and usability of \textit{SCSimulator}.

\subsection{Participants}
\par We recruited 12 new participants (\textbf{P1-P12}) from research institutes and companies at partner universities (5 female, 7 male; aged 23 - 40). All participants had prior corporate internship experience, with most having contributed to academic publications. While they all had experience using LLMs in professional or research contexts, none had prior experience using LLM-driven MAS. To explore potential differences between academic and industry perspectives, we divided the participants into two groups, \textit{\textbf{Research}} and \textit{\textbf{Enterprise}}. Notably, \textbf{P1} was assigned to the \textit{\textbf{Research}} group, as he is a professor of Management, while \textbf{P7-P8} was assigned to the \textit{\textbf{Enterprise}} group, as they are managers in companies. The remaining participants, primarily students, were assigned randomly. This grouping approach was chosen because many participants had overlapping academic and industrial experience, making background-based classification challenging. Thus, the primary distinction between the groups was based on task context, with each simulating different professional roles.

\begin{table}[h]
\centering
\caption{Demographic Information of Participants.}
\label{tab:user_study_participants}
\begin{tabular}{ccccccc}
\hline
\textbf{ID} & \textbf{Gender} & \textbf{Age} & \textbf{Education} & \textbf{Group} \\ \hline
P1 & Male & 35 & Ph.D. &  \\
P2 & Female & 30 & Ph.D. &  \\
P3 & Male & 28  & Ph.D. &  \\
P4 & Female & 24 & Master &  \\
P5 & Male & 27 & Master  &  \\
P6 & Female & 23 & Master & \multirow{-6}{*}{\textit{\textbf{Research}}} \\
\hline
P7 & Female & 40 & Master &  \\
P8 & Male & 35 & Master &  \\
P9 & Male & 30 & Ph.D. &   \\
P10 & Male & 24 & Master &   \\
P11 & Male & 29 & Ph.D. &   \\
P12 & Female & 26 & Master &  \multirow{-6}{*}{\textit{\textbf{Enterprise}}} \\
\hline
\end{tabular}
\end{table}

\subsection{Study Setting \& Procedure}

\par \textbf{Preparation.} We reused the dataset described in \autoref{sec:technical_evaluation} to ensure consistency across study conditions. 

\par \textbf{Introduction (15 min).} Participants received an overview of the study goals and procedures, completed a brief demographic questionnaire, and provided consent for interaction logging during the study. The session concluded with a guided demonstration of the core features of \textit{SCSimulator} using an example scenario.

\begin{table*}[t]
\centering
\renewcommand{\arraystretch}{1.15}
\begin{tabular}{p{0.46\linewidth} p{0.46\linewidth}}
\toprule
\textit{\textbf{Research}} &
\textit{\textbf{Enterprise}} \\
\midrule
You will take the perspective of an \textit{industry analyst} who studies the structural evolution of the paper packaging sector. This industry has long been characterized by ibstable SCs dominated by a few traditional manufacturers that prioritize cost control and operational stability. In recent years, however, several emerging firms have introduced more sustainable and transparent production models, gradually reshaping the competitive and collaborative landscape. 

Your task is to use \textit{SCSimulator} to explore how these shifts influence the overall SC network. Focus on identifying patterns of competition, collaboration, and adaptation among traditional and emerging players. Pay attention to changes in operational relationships, technological diffusion, and reputation dynamics across the network. As you explore, please explain your reasoning and decisions aloud.

&
You will take the perspective of a \textit{manager at enterprise 4838}, a long-established leader in the paper product industry. Your company has traditionally succeeded by maintaining cost efficiency and reliable supply, but the market landscape is changing. New competitors like \textbf{7437} and \textbf{6174} are gaining traction through innovative and environmentally friendly manufacturing approaches. 

Your task is to use \textit{SCSimulator} to analyze potential new customer and partner opportunities for \textbf{4838}. Consider whether it would be more beneficial to collaborate with or compete against these emerging firms. Evaluate each option in terms of operational stability, technological advantage, and market reputation. As you explore, please explain your reasoning and decisions aloud. \\
\bottomrule
\end{tabular}
\caption{Task Description for Two Study Groups.}
\label{tab:taskDescription}
\end{table*}
\par \textbf{Tasked-based Simulation (80 min).} In this stage, participants were introduced to the SC dataset and assigned an analysis task with varying descriptions based on their group (refer to \autoref{tab:taskDescription}). They completed the task using a think-aloud approach, verbalizing their reasoning and referencing relevant evidence observed in \textit{SCSimulator}.

\par \textbf{Feedback (30 min).} After completing the task, participants filled out a 5-point Likert-scale questionnaire evaluating the cognitive workload (Q1-Q6), system usability (Q7-Q12) and effectiveness (Q13-Q17) of \textit{SCSimulator}. Each item was rated from 1 (strongly disagree) to 5 (strongly agree). Additional open-ended questions (Q18–Q20) explored potential applications of LLM-driven MAS in participants' workflows and gathered feedback on system limitations and areas for improvement. Participants were encouraged to provide detailed verbal explanations to contextualize their ratings and insights.

\subsection{Results and Analysis}
\par \autoref{tab:user_study_results} summarizes participants' responses to the feedback questionnaire. White cells represent the \textbf{\textit{Research}} group, while gray cells represent the \textbf{\textit{Enterprise}} group. Overall, \textit{SCSimulator} received positive ratings, with most scores for the closed-ended questions being favorable. All participants agreed that the system facilitated the exploration of partner selection within SC using LLM-driven MAS. However, the scores for mental demand, temporal demand, and ease of learning were less favorable. Detailed feedback is provided below.

\begin{table*}[h]
\centering
\caption{Feedback questionnaire results. Questions Q1-Q17 are closed-ended and rated on a 5-point Likert scale. Questions Q18-Q20 are open-ended for participant feedback. Data from the \textbf{\textit{Research}} group are shown in white cells, while the \textbf{\textit{Enterprise}} group data are in gray cells.}
\label{tab:user_study_results}
\begin{tabular}{ccccc}
\hline
Categories & Factors & Question & Mean & S.D. \\ \hline
\multirow{12}{*}{NASA-TLX} & \multirow{2}{*}{\begin{tabular}[c]{@{}c@{}}Mental\\ Demand\end{tabular}} & \multirow{2}{*}{\begin{tabular}[c]{@{}c@{}}Q1: How mentally demanding\\ was the task?\end{tabular}} & 2.16 & 0.98 \\
 &  &  & \cellcolor{gray!20}2.33 & \cellcolor{gray!20}0.52 \\ \cline{2-5} 
 & \multirow{2}{*}{\begin{tabular}[c]{@{}c@{}}Physical\\ Demand\end{tabular}} & \multirow{2}{*}{\begin{tabular}[c]{@{}c@{}}Q2: How physically demanding\\  was the task?\end{tabular}} & 1.5 & 0.55 \\
 &  &  & \cellcolor{gray!20}1.33 & \cellcolor{gray!20}0.52 \\ \cline{2-5} 
 & \multirow{2}{*}{\begin{tabular}[c]{@{}c@{}}Temporal\\ Demand\end{tabular}} & \multirow{2}{*}{\begin{tabular}[c]{@{}c@{}}Q3: How hurried or rushed \\ was the pace of the task?\end{tabular}} & 2.33 & 0.52 \\
 &  &  & \cellcolor{gray!20}2.5 & \cellcolor{gray!20}0.55 \\ \cline{2-5} 
 & \multirow{2}{*}{Performance} & \multirow{2}{*}{\begin{tabular}[c]{@{}c@{}}Q4: How successful were you in accomplishing\\ what you were asked to do?\end{tabular}} & 4.17 & 0.41 \\
 &  &  & \cellcolor{gray!20}4.33 & \cellcolor{gray!20}0.52 \\ \cline{2-5} 
 & \multirow{2}{*}{Effort} & \multirow{2}{*}{\begin{tabular}[c]{@{}c@{}}Q5: How hard did you have to work\\ to accomplish your level of performance?\end{tabular}} & 2.83 & 0.75 \\
 &  &  & \cellcolor{gray!20}2.67 & \cellcolor{gray!20}0.52 \\ \cline{2-5} 
 & \multirow{2}{*}{Frustration} & \multirow{2}{*}{\begin{tabular}[c]{@{}c@{}}Q6: How insecure, discouraged, irritated,\\ stressed, and annoyed were you?\end{tabular}} & 1.67 & 0.52 \\
 &  &  & \cellcolor{gray!20}1.83 & \cellcolor{gray!20}0.41 \\ \hline
\multirow{12}{*}{Usability} & \multirow{2}{*}{\begin{tabular}[c]{@{}c@{}}Easy\\ to Use\end{tabular}} & \multirow{2}{*}{Q7: I thought the system was easy to use.} & 3.83 & 0.41 \\
 &  &  & \cellcolor{gray!20}4.17 & \cellcolor{gray!20}0.41 \\ \cline{2-5} 
 & \multirow{2}{*}{Functions} & \multirow{2}{*}{\begin{tabular}[c]{@{}c@{}}Q8: I found the various functions \\ in this system were well integrated.\end{tabular}} & 4.5 & 0.55 \\
 &  &  & \cellcolor{gray!20}4.33 & \cellcolor{gray!20}0.52 \\ \cline{2-5} 
 & \multirow{2}{*}{\begin{tabular}[c]{@{}c@{}}Quick\\ to Learn\end{tabular}} & \multirow{2}{*}{\begin{tabular}[c]{@{}c@{}}Q9: I think most people would \\ learn to use this system very quickly.\end{tabular}} & 1.67 & 0.82 \\
 &  &  & \cellcolor{gray!20}1.33 & \cellcolor{gray!20}1.03 \\ \cline{2-5} 
 & \multirow{2}{*}{Frequency} & \multirow{2}{*}{\begin{tabular}[c]{@{}c@{}}Q10: I think I would like to \\ use this system frequently.\end{tabular}} & 4.67 & 0.52 \\
 &  &  & \cellcolor{gray!20}4.33 & \cellcolor{gray!20}0.52 \\ \cline{2-5} 
 & \multirow{2}{*}{Confidence} & \multirow{2}{*}{Q11: I felt very confident using the system.} & 4.17 & 0.41 \\
 &  &  & \cellcolor{gray!20}4.17 & \cellcolor{gray!20}0.41 \\ \cline{2-5} 
 & \multirow{2}{*}{Inconsistency} & \multirow{2}{*}{\begin{tabular}[c]{@{}c@{}}Q12: I thought there was too much\\  inconsistency in this system.\end{tabular}} & 1.50 & 0.55 \\
 &  &  & \cellcolor{gray!20}1.67 & \cellcolor{gray!20}0.52 \\  \hline
\multirow{10}{*}{Effectiveness} & \multirow{2}{*}{\begin{tabular}[c]{@{}c@{}}The \textit{Upload Page} \&\\ the \textit{Control Panel}\end{tabular}} & \multirow{2}{*}{\begin{tabular}[c]{@{}c@{}}Q13: The \textit{Upload Page} and the \textit{Control Panel}\\  help initialize MAS.\end{tabular}} & 4.67 & 0.52 \\
 &  &  & \cellcolor{gray!20}4.33 & \cellcolor{gray!20}0.52 \\ \cline{2-5} 
 & \multirow{2}{*}{\begin{tabular}[c]{@{}c@{}}Global\\ View\end{tabular}} & \multirow{2}{*}{\begin{tabular}[c]{@{}c@{}}Q14: The \textit{Global View} shows\\  the overall simulation evolution.\end{tabular}} & 4.33 & 0.82 \\
 &  &  & \cellcolor{gray!20}4.33 & \cellcolor{gray!20}0.52 \\ \cline{2-5} 
 & \multirow{2}{*}{\begin{tabular}[c]{@{}c@{}}Focus\\ View\end{tabular}} & \multirow{2}{*}{\begin{tabular}[c]{@{}c@{}}Q15: The \textit{Focus View} shows the network \\ dynamics impact on target agents.\end{tabular}} & 4.17 & 0.75 \\
 &  &  & \cellcolor{gray!20}4.5 & \cellcolor{gray!20}0.55 \\ \cline{2-5} 
 & \multirow{2}{*}{\begin{tabular}[c]{@{}c@{}}Adjustment\\ View\end{tabular}} & \multirow{2}{*}{\begin{tabular}[c]{@{}c@{}}Q16: The \textit{Adjustment View} shows agents' decision\\ procedure and allow user edits\end{tabular}} & 4.5 & 0.55 \\
 &  &  & \cellcolor{gray!20}4.17 & \cellcolor{gray!20}0.75 \\ \cline{2-5} 
 & \multirow{2}{*}{\begin{tabular}[c]{@{}c@{}}Simulation\\ Path View\end{tabular}} & \multirow{2}{*}{\begin{tabular}[c]{@{}c@{}}Q17: The \textit{Simulation Path View} presents\\ the user's simulation journey.\end{tabular}} & 4.17 & 0.75 \\
 &  &  & \cellcolor{gray!20}4.17 & \cellcolor{gray!20}0.75 \\ \hline
\multirow{3}{*}{Open-ended} & / & \multicolumn{3}{c}{Q18: How do you think SCSimulator might influence your current workflow?} \\
\cline{2-5} 
& / & \multicolumn{3}{c}{Q19: What are the advantages and disadvantages of SCSimulator?} \\ \cline{2-5} 
 & / & \multicolumn{3}{c}{Q20: How can SCSimulator be improved in the future?} \\ \hline
\end{tabular}
\end{table*}

\subsubsection{Cognitive Workload}
\par Cognitive workload was assessed using the NASA-TLX scale, covering mental, physical, and temporal demands, as well as performance, effort, and frustration. As shown in \autoref{tab:user_study_results}, participants reported low physical demand and frustration, with high performance scores across both groups. In contrast, mental demand and effort were rated higher, reflecting the complexity and interactivity of LLM-driven MAS. Temporal demand was moderate, likely due to real-time simulation delays, which sometimes caused brief interruptions or waiting periods.

\subsubsection{Usability}
\par The usability of \textit{SCSimulator} was evaluated across several factors, including ease of use, functionality, learnability, frequency of use, user confidence, and perceived consistency. As shown in \autoref{tab:user_study_results}, \textit{SCSimulator} received high scores in overall usability. Participants found the system well-integrated and easy to use once they became familiar with it. However, ratings for learnability were relatively lower. Many users found some of the visual designs challenging to understand and required additional guidance from the facilitators, particularly regarding the \textit{Focus View}. Despite this, participants generally considered the interface to be coherent and reliable after some practice.


\subsubsection{Effectiveness}
\par Based on responses to Q13–Q17, all participants endorsed the framework and visual design of \textit{SCSimulator}. They confirmed that the system facilitated efficient initialization and monitoring of LLM-driven MAS simulations. As \textbf{P1} noted, ``\textit{I never thought it would be this easy to run a [LLM-driven multi-agent] simulation. Previously, MAS in code form required a lot of time for coding and debugging... It is easier to use SCSimulator.}''

\par Participants also expressed strong appreciation for our integration of traditional modeling techniques with the qualitative CoT reasoning generated by agents. This hybrid approach effectively alleviates common concerns regarding hallucinations in LLM-based systems. As \textbf{P10} noted, ``\textit{SHAP values provide me with a quantitative anchor that clearly indicates which attributes truly drive the simulation outcomes. I then use the agents’ explanation to examine whether their reasoning aligns with these weights. This form of cross-validation makes the results far more credible than presenting a final decision alone.}''. In addition, participants praised our berry-based metaphor for its ability to reveal interactions between suppliers and customers, despite its initial complexity. As \textbf{P9} observed, ``\textit{Although it is hard to understand at first glance, once I become familiar with it, I can clearly see how we (the focal company) are influenced by upstream and downstream partners.}''

\par The combination of the \textit{Global View} and \textit{Focus View} provided both an overview and the ability for detailed exploration, helping to mitigate the data overload commonly associated with traditional dashboards. As \textbf{P3} noted, ``\textit{Unlike in daily work where multiple tabs are needed to compare firms, \textit{SCSimulator} allowed faster and more integrated inspection.}'' Similarly, the \textit{Adjustment View}'s expandable tree design helped participants manage large volumes of simulation data without confusion.

\par Notably, some participants observed that agents exhibited negotiation-like behaviors, such as offering customized services. \textbf{P5} commented, ``\textit{It makes me feel like I am observing a real company's strategic game.}'' However, as the simulation does not support fine-grained operations, some participants raised concerns about agents acting beyond realistic boundaries. \textbf{P12} stated, ``\textit{If an agent can cooperate through cost-free deception, it undermines the realism of the simulation.}'' This phenomenon will be discussed further in the Discussion section (\autoref{sec:dis_maa}).

\section{Discussion}
\par This section situates \textit{SCSimulator} within the broader landscape of SC modeling and decision support. We synthesize how different user constituencies, researchers and enterprise practitioners, perceive and leverage the system, and we discuss the principal design implications, constraints on scalability and validation, and directions for future work.

\subsection{Expanding Knowledge Boundaries: From Rapid Prototyping to Cognitive Augmentation}

\subsubsection{\textit{SCSimulator} as an Exploratory Medium Rather than a Predictive Engine}

\par The central design rationale of \textit{SCSimulator} is to support exploratory sensemaking by enabling users to examine plausible yet unrealized SC scenarios that are often obscured by real-world complexity. Rather than optimizing for predictive accuracy, the system foregrounds MAS-supported “what-if” exploration. This form of exploration is intended to improve decision quality not by producing optimal answers, but by expanding the space of considered alternatives and making implicit assumptions explicit. This emphasis distinguishes \textit{SCSimulator} from traditional graph-based models, mathematical formulations, and visual analytics systems, and reflects an explicit trade-off: analytical rigor is relaxed in favor of expressive flexibility. As a result, users can probe counterfactual configurations and alternative futures that are difficult to enumerate or formalize \textit{a priori}.

\par For researchers, \textit{SCSimulator} functions as an experimental sandbox for hypothesis exploration. By manipulating environmental assumptions and agent behaviors through global domain knowledge, researchers can observe how LLM-driven agents extrapolate system dynamics along diverse trajectories. While these outcomes are not statistically grounded, they surface qualitative patterns and potential mechanisms that are often inaccessible to rule-based or optimization-driven simulations, whose behavioral spaces are comparatively constrained.

\par In contrast, enterprise participants primarily interpret \textit{SCSimulator} as a decision exploration tool rather than a mechanism-oriented simulator. They tend to issue concrete behavioral instructions to individual agents and examine the resulting interactions to assess possible operational consequences. This interaction style allows users to temporarily adopt counterpart perspectives within the SC, reducing reliance on intuition-driven or single-perspective reasoning and mitigating subjective biases during strategic deliberation. As a result, decision quality is supported through improved perspective-taking, more balanced consideration of trade-offs, and reduced overreliance on heuristic judgment.

\par These divergent orientations also shape how simulation outputs are interpreted. Researchers typically treat generated outcomes as plausible but non-objective distributions, suitable for hypothesis generation rather than empirical inference. Enterprise users, by comparison, engage with simulations as narrative-driven scenario instantiations that support reflective decision-making under uncertainty. While \textit{SCSimulator} accommodates both research- and enterprise-oriented use cases, its current support is primarily limited to early-stage exploration. For example, enterprise users expressed interest in finer-grained interactions, such as participating in multi-round business negotiations rather than issuing high-level behavioral directives. Future work will therefore investigate scenario-specific interfaces and analytical affordances that more fully leverage LLM-driven MAS for both experimental inquiry and operational reasoning.

\subsubsection{Scalability and the Transition to Quantitative Validation}

\par The current implementation of \textit{SCSimulator} typically operates with 30 to 50 agents. While the precise number of firms required to reach \textit{theoretical saturation}\footnote{Theoretical saturation refers to the point in qualitative research at which additional data collection yields no new conceptual insights and all thematic categories are sufficiently developed.}~\cite{rahimi2024saturation} in SC modeling remains an open question, prior work~\cite{rwakira2015supply, lewis1998iterative, eisenhardt1989building, eisenhardt2007theory} suggests that this scale, 30 to 50 agents, provides a reasonable foundation for exploratory qualitative analysis. Moreover, empirical studies by Fujiwara et al.~\cite{fujiwara2010large} indicate that the average firm maintains approximately four upstream and downstream partners, suggesting that \textit{SCSimulator} is well aligned with the structural characteristics of most manufacturing firms.

\par Nevertheless, this scale is insufficient for rigorous quantitative validation, a limitation emphasized by our research participants. Fujiwara et al.~\cite{fujiwara2010large} also document a pronounced long-tail effect in SC connectivity, wherein a small number of hub firms maintain relationships on the order of $O(10^2)$. Modeling such enterprises would require substantially different computational architectures and interface designs to manage the resulting complexity. In addition, expert interviews highlighted concerns regarding the controllability of agent behavior. This form of ``behavioral entropy'' is likely to increase rapidly as simulation scale grows, undermining the reliability of system-level outcomes when hundreds of autonomous agents are involved. Addressing this issue will require more sophisticated mechanisms for constraining and coordinating agent autonomy (see \autoref{sec:dis_maa}).

\par In summary, while \textit{SCSimulator} demonstrates clear potential for qualitative validation, consistent with its role as a heuristic and exploratory system, scaling toward quantitative robustness remains a central challenge and an important direction for future research.

\subsubsection{Real-World Deployment Considerations and Practical Constraints}

\par While the preceding subsections discuss the cognitive role of \textit{SCSimulator} and its scalability limits from an analytical perspective, this subsection reflects on broader constraints that shape real-world deployment. In industrial contexts, adoption is influenced not only by system capability, but also by data availability, cost–benefit trade-offs, and the reliability and safety of model-generated behaviors.

\par \textbf{Data availability and knowledge substitution.}
In SC partner selection scenarios, data are inherently incomplete and fragmented. Firm-level decision rationales, informal collaborations, and strategic intentions are rarely observable, which limits the applicability of data-intensive modeling approaches, like deep learning. In addition, effective mathematical models often require deep, long-term domain expertise, resulting in high organizational cost.

Rather than treating data as a basis for precise fitting, \textit{SCSimulator} uses data as contextual anchors for plausible behavioral reasoning. LLM-driven agents are intended to approximate abstract decision logics of enterprises rather than reproduce observed outcomes. Domain experts can directly shape agent decision styles through explicit \textit{Knowledge} inputs, partially compensating for missing or unobservable data by externalizing tacit expertise that is difficult to formalize. Nevertheless, LLM-driven MAS do not resolve information asymmetries inherent in real-world SCs; they primarily mitigate these asymmetries by making assumptions explicit and supporting reflective exploration.

\par \textbf{Cost, cognitive burden, and diminishing returns of scale.}
Although large-scale MAS~\cite{pan2024very, saravanos2023distributed,guo2025mass} has become increasingly feasible from a technical and economic perspective, scalability alone does not determine practical viability. In interactive analytical settings, cognitive burden and interpretability impose equally strong constraints. As the number of agents increases, MAS generate large volumes of unstructured interactions that are difficult to synthesize and reason about.

While \textit{SCSimulator} maintains manageable cognitive load at small to medium scales, it remains unclear whether similar interpretability can be preserved at much larger scales. Besides, from the perspective of theoretical saturation, early simulation rounds and moderate agent populations are more likely to yield novel insights, whereas further expansion often leads to diminishing returns. For users seeking qualitative sensemaking, small-scale simulations with prioritized early rounds may therefore offer the most favorable cost–benefit trade-off.

\par \textbf{Reliability, safety, and expert oversight.}
In deployment scenarios, the behavioral boundaries of autonomous agents remain inherently ambiguous, particularly when agents are driven by generative models. Without appropriate constraints, LLM-based agents may exhibit unstable or contextually inappropriate reasoning, posing risks in sensitive decision-making domains.

Accordingly, \textit{SCSimulator} is not positioned as an autonomous decision-making system. Expert oversight is treated as a necessary condition for reliable use, providing a mechanism for validating assumptions, constraining agent behavior, and correcting undesirable trajectories. Reliability thus emerges from continuous human engagement rather than full automation, reinforcing the role of LLM-driven MAS as tools for structured reflection and scenario exploration rather than substitutes for expert judgment.

\subsection{Design Implication}

\subsubsection{Explainable Multi-Agent Reasoning}
\par Interpretability is crucial fostering users' trust and validating insights. While LLMs can provide natural language explanations, such as CoT, prior studies have highlighted that these narratives are often unreliable and difficult to verify~\cite{turpin2023language}. In the context of partner selection within SC, relying solely on textual reasoning does not adequately capture the quantitative impact of agent decisions. \textit{SCSimulator} bridges this gap by integrating both qualitative and quantitative perspectives. Specifically, CoT generated by LLMs reveals the internal rationale behind decisions, while predictive models estimate agent performance, allowing for a more explicit understanding of network dynamics. This multi-faceted interpretability allows users to connect localized reasoning with broader system-level effects, thereby enhancing both comprehension and trust in simulation outcomes.

\subsubsection{Visual Design for Reducing Cognitive Overload}
\par LLM-driven MAS simulations can generate large volumes of low-value logs, which may overwhelm users. Visualization plays a key role in reducing cognitive load~\cite{huang2009measuring, zhu2010visualization}. In \textit{SCSimulator}, we adopt the ``\textit{overview first, zoom and filter, then details-on-demand}'' design principle. The \textit{Global View} offers a comprehensive view of the entire simulation, allowing users to identify high-value nodes. In the \textit{Focus View}, an egocentric network visualization highlights the specific influence of these nodes on network dynamics. The \textit{Adjustment View} organizes textual simulation data in an expandable tree structure. This multi-level design approach effectively mitigates cognitive overload while supporting targeted exploration and intervention.

\subsubsection{Managing Agent Autonomy}
\label{sec:dis_maa}
\par User preferences regarding agent autonomy vary significantly. Users who favor high-autonomy (\textbf{P1, P3, P4, P5}) appreciate agents' ability to exceed preset expectations and generate insights through unexpected actions. As \textbf{P1} noted, ``\textit{Simulation is meaningful precisely under uncertainty.}'' An example of this is when agents selectively adhere to knowledge, especially when \textit{Knowledge} input by users are unsuitable for the current context. On the other hand, supporters of low autonomy (\textbf{P2, P7, P12}) expressed concerns that unrestricted agents may exploit rule boundaries, leading to unrealistic outcomes, such as unfulfillable promises. These findings highlight the importance of carefully managing agent freedom in LLM-MAS systems.

\par Two key design implications arise from these observations. First, defining the level of agent autonomy is critical. Prior work categorizes autonomy levels, ranging from operator-controlled to fully observational~\cite{feng2025levelsautonomyaiagents}, but establishing precise behavioral boundaries remain an open challenge, particularly given the diverse preferences of users. Second, managing boundary violations requires thoughtful design. Rather than simply halting execution, nudging agents back into acceptable behaviors may offer a more flexible alternative~\cite{ma2025agentdynexnudgingmechanicsdynamics}. Overall, careful consideration of the agent's decision-making scope is essential when designing LLM-driven MAS.

\subsection{Generalizability}
\par Although our work primarily focuses on partner selection within SC, the underlying mechanism can be framed as edge formation and deletion in social networks. Similar dynamics can be explored in other domains, such as logistics, transportation systems, or multi-agent communities. By simply adjusting the prompt within \textit{SCSimulator}, users can easily adapt the framework to these new contexts. Additionally, through a user study with participants from diverse backgrounds, the \textit{SCSimulator} framework has demonstrated strong generalizability. Developers can build upon this framework by adhering to a human-in-the-loop approach and using a \textit{simulate}–\textit{check}–\textit{adjust} cyclic process to tailor it for specific tasks. Furthermore, our egocentric network visualization, inspired by the ``berry'' metaphor, facilitates detailed interpretation of ego–alter interactions. This design is not limited to LLM-driven MAS and can be directly applied to other SC visualization contexts, provided the task involves analyzing how the structural properties of nodes influence their performance or behavior. More broadly, social network applications with similar analytical objectives can also leverage this metaphor by adapting the visual encoding accordingly.

\subsection{Limitation and Future Work}

\subsubsection{Study Design}

\par While our evaluation indicates high usability and perceived practical value, it was conducted in short-term, laboratory-based settings. Many SC decisions unfold over extended periods, making long-term effects difficult to capture in such studies. 
Moreover, we did not include direct comparisons with baseline systems in our user study, primarily due to the lack of suitable baselines for fair comparison. MAS–based approaches differ substantially from mainstream mathematical optimization models or expert-driven analytical tools in their underlying assumptions, interaction mechanisms, and modes of sensemaking. In practice, \textit{SCSimulator} is intended to complement rather than replace these established methods. In addition, in a highly domain-specific context such as SCM, applying a generic MAS framework, particularly without careful consideration of interactive and visual design, would be inappropriate.
The study also relied on a relatively small, domain-specific dataset, which limits generalizability. This constraint stems from the scarcity of high-quality SC data \cite{hazen2014data} and from our decision to construct a focused dataset in close collaboration with domain experts from the paper products sector. Future work will prioritize long-term, in-situ deployments of \textit{SCSimulator} within partner organizations to examine sustained use, downstream decision impacts, and more ecologically valid evaluations, including comparisons grounded in real operational practices. We also plan to expand data collection and evaluate the framework on larger and more diverse SC datasets.

\subsubsection{System Design}
\par Currently, \textit{SCSimulator} has limited support for large-scale MAS, involving more than 50 agents, due to the performance constraints of API-based LLM access. Existing work on small-sized LLMs~\cite{belcak2025smalllanguagemodelsfuture} and large-scale LLM-driven MAS frameworks~\cite{pan2024very} points to promising directions for improvement. Future work will explore integrating these techniques to enable simulations with a larger number of agents. Additionally, the MAS in \textit{SCSimulator} currently supports only basic commercial models. Observed negotiation behaviors from LLM agents indicate the potential to simulate more complex business interactions. Future enhancements will focus on extending \textit{SCSimulator} with more advanced mechanisms, such as bidding processes, and exploring dynamics under hybrid models. Finally, our current LLM relies on persona-based prompts, which may lack specialized SC expertise. By integrating knowledge bases and employing retrieval-augmented generation (RAG), we can provide more domain-specific capabilities. We plan to collaborate with experts to develop specialized LLMs, enhancing reasoning and decision-making within LLM-driven MAS.


\section{Conclusion}

\par In this study, we introduce \textit{SCSimulator}, an interactive framework designed to leverage LLM-driven MAS for partner selection within SCs. Collaborating with five domain experts, we identified key challenges in their current workflows, motivating us to incorporate LLM-driven MAS. Through co-design with experts, we gained three findings into expert interactions with LLM-driven MAS. Seven concrete design requirements were extracted from the two-phase formative study. Guided by these insights, \textit{SCSimulator} provides an interactive visual analytics environment where users can monitor supplier–customer dynamics, interpret agent decisions via integrated CoT reasoning, predictive models, and SHAP, and strategically intervene to align simulations with their expectations. A usage scenario and a user study ($N=12$) demonstrated the framework's effectiveness and usability, highlighting its potential as a rapid prototyping tool that enables users to quickly run simulations, validate ideas, and gain actionable insights.

\section{GenAI Usage Disclosure}
\par In this work, GenAI tools were used to assist in several stages of research and writing process. During the development phase, code generation tools such as GitHub Copilot were employed to accelerate implementation. For manuscript preparation, LLMs, like ChatGPT and Gemini, were used to refine wording, improve grammar, and enhance clarity. All AI-generated content was carefully reviewed and edited by the authors to ensure accuracy and precision. Besides, one of figure used to illustrate our metaphor are generated by Nano Banana. In addition, as this study focuses on LLM-driven MAS, the simulation results were generated through interactions among LLM-driven agents tailored to our scenario of partner selection within SC.

\begin{acks}
We gratefully acknowledge the anonymous reviewers for their insightful feedback. This research was supported by the National Natural Science Foundation of China (No. 62372298), the National Key Research and Development Program of China under Grant 2022YFF0903302, the Shanghai Engineering Research Center of Intelligent Vision and Imaging, the Shanghai Frontiers Science Center of Human-centered Artificial Intelligence (ShangHAI), and the MoE Key Laboratory of Intelligent Perception and Human-Machine Collaboration (KLIP-HuMaCo).
\end{acks}

\bibliographystyle{ACM-Reference-Format}
\bibliography{sample-base}

\newpage
\appendix

\section{Prompts}

\label{app:prompt}

\begin{lstlisting}[caption={CompanyInfo2Prompt}, label={Company Information}]
You are company {company_id}.
Your industry is {company_industry[company_id]}.
The following is global knowledge: {global_knowledge}
The following is your company-specific knowledge: {company_knowledge}.
At {time_stamp}, your feature info is: {feature_info_self}
You have the following supply chain connections (supplier, customer):
supplier: {network_iter['supplier']}
{for supplier_id in network_iter['supplier']}: 
    Supplier {supplier_id} industry is {company_industry[supplier_id]}. feature info: {feature_info_supplier}
customer: {network_iter['customer']}
{for customer_id in network_iter['customer']}: 
    Customer {customer_id} industry is {company_industry[customer_id]}. feature info: {feature_info_customer}
\end{lstlisting}

\begin{lstlisting}[caption={Prompt of Stage I —— Plan}, label={stage1}]
You are now an company manager and you are making decisions on the management of your supply chain in next timestampe.
The user will provide supply chain network and company information within this supply chain.
You need to fully analyze the information provided by the user and propose plans. 
A plan should include you intention, you reason with step-by-step thinking, and whether you are plan to increase or decrease partners. 
If true as you plan to increase, you should provide query requirement for company query.
You response should fully following the following content in strict JSON format:
```json
[
    {
        "plan": "your plan' brief description",
        "reason": "your reason for this Plan"
        "is_seek_collaboration": True/False "whether you plan to increase collaboration or decrease"
        "is_seek_suppliers": True/False //if true, you are seek more suppliers otherwise you are seek more customers. Only work if is_seek_collaboration=True
    },
    ...
]

The following is information about your company and its supply chain partners:
{company_info2prompt(...)}


\end{lstlisting}

\begin{lstlisting}[caption={Prompt of Stage II —— Query}, label={stage2}]
You are now an company manager and you are making decisions on the management of your supply chain in next timestampe.
The user will provide supply chain network and company information within this supply chain.
You should fully analyze the information provided by the user and propose plans. Then, you should list the constrain of industry and the weighted score of each features. Those information will be used for querying potential company.
You should response with the same number of plans as the user provided.
You response should fully following the following content in strict JSON format:
```json
[
    {
        "industry_set": [industry1, ...],//the constrain on company industries, you may return a empty list to ignore this constrain
        "weighted_scores": [{ feature: xxx, weight:xxx },{ feature: xxx, weight:xxx }] ,//weighted score of each feature,only output in feature_col_list: [...]
    },
    ...
]

The following is information about your company and its supply chain partners:
{company_info2prompt(...)}

The following is your plan list:
{for i, plan_iter in enumerate(plan):}
    Plan {i+1}: {plan_iter['plan']}. With reason: {plan_iter['reason']}.


\end{lstlisting}

\begin{lstlisting}[caption={Prompt of Stage III —— Request}, label={stage3}]
You are now an company manager and you are making decisions on the management of your supply chain in next timestampe.
The user will provide supply chain information and plan list with potential candidate if he want to increase collaboration.
You should fully analyze the information provided by the user and propose plan. Then, you should make detail decisions. Specifically, if the "is_added" is false, you should choose one or some company and cancel collaborations with them with step by step thinking. If ths "is_added" is true, you should check the candidate and determine one or some company and request collaborations with them,  with step by step thinking.
You answer shoule include the list of id that you decided to cancel or request collaboration for the plan. For a plan that does not want to increase collaborators, you should choose companies from your existing supply chains to cancel. The outer dimension is plan and the inner is id
You should strictly follow the format below.
```json
    [
    [{
          "company_id": "the id of company",
      "is_chosen": True/False//  whether you determine to choose this candidate to build collaboration and output even if false.
        "reason": "the reason why you make such a decision",
        "extra_info": "The information that would be sent to the company to facilitate collaboration. Only works when this plan is a seek collaboration plan and is_chosen is True.
    },
    ...],
    ...]

The following is information about your company and its supply chain partners.
{company_info2prompt(...)}

The following is information about your company and its supply chain partners.
{company_info2prompt(...)}

The following is your plan list with potential companies:
{for i in range(len(plan)):}
    Plan {i+1}: {plan_iter['plan']}. With reason: {plan_iter['reason']}.
    Your query constrain is industry in {query_iter['industry_set']} and feature weighted score is {query_iter['weighted_scores']}.
    if {plan_iter["is_seek_collaboration"]}:
        The following is your candidate company list:
        ...
\end{lstlisting}

\begin{lstlisting}[caption={Prompt of Stage IV —— Reply}, label={stage4}]
You are now an company manager and you are making decisions on the management of your supply chain in next timestampe.
The user will provide supply chain information and collaboration requests from other companies. You have to decide whether collaborate or not for each request one by one and give corresponding reasoning with step-by-step thinking.
You response and fully follow the following content in strict JSON format:
```json
[
    {
      "company_id": "the id of the company requesting",
      "is_accepted": True/False, // Do you accept this collaboration 
      "reason": "the reason"
    },
    ...
]

The following is information about your company and its supply chain partners.
{company_info2prompt(...)}

The following is the collaboration requests you received:
Request from company {receive_requests['wanna_to_be_supplier']} to be your supplier.
Request from company {receive_requests['wanna_to_be_customer']} to be your customer.
\end{lstlisting}


\end{document}